\documentclass[aps,prb,twocolumn,showpacs,superscriptaddress]{revtex4-1}
\usepackage{latexsym}
\usepackage{amssymb}
\usepackage{graphicx}
\usepackage{amsmath}
\usepackage{tikz}
\usepackage{bm}
\usepackage[colorlinks,
linkcolor=black,
citecolor=black,
urlcolor=blue
]{hyperref}
\usepackage{verbatim}
\usepackage{mathrsfs}
\usepackage{extarrows}
\usepackage{comment}
\usepackage{mathtools,slashed}
\usepackage{soul}
\usepackage[toc,page]{appendix}
\usepackage[vcentermath]{youngtab}
\usepackage{multirow}
\usepackage[width=.75\textwidth]{caption}
\usepackage{atbegshi,picture}
\usepackage{lipsum}

\makeatletter
\renewcommand\@makecaption[2]{
	\par
	\vskip\abovecaptionskip
	\begingroup
	\small\rmfamily
	\begingroup
	\samepage
	\flushing
	\let\footnote\@footnotemark@gobble
	\@make@capt@title{#1}{#2}\par
	\endgroup
	\endgroup
	\vskip\belowcaptionskip
}
\makeatother

\AtBeginShipoutNext{\AtBeginShipoutUpperLeft{
		\put(\dimexpr\paperwidth-1cm\relax,-1.5cm){\makebox[0pt][r]}
}}

\makeatletter\begin{document}
\title{Duality viewpoint of criticality}

\author{Linhao Li}

\affiliation{Institute for Solid State Physics, The University of Tokyo. Kashiwa, Chiba 277-8581, Japan}

\author{Yuan Yao}
\thanks{yuan.yao@riken.jp}
\affiliation{Condensed Matter Theory Laboratory, RIKEN CPR, Wako, Saitama 351-0198, Japan}

\begin{abstract}
In this work, we study quantum many-body systems which are self-dual under duality transformation connecting different symmetry protected topological (SPT) phases. We provide a geometric explanation of the criticality of these self-dual models. More precisely, we show a ground state (quasi-)degeneracy under the periodic boundary conditions,i.e., the ingappability of the bulk spectrum. Equivalently, the symmetry group at criticality, including the duality symmetry, has a mixed 't Hooft anomaly. This approach can not only predict the spectrum of the self-dual model with ordinary 0-form symmetry, but also be applied to that with generalized symmetry, such as higher form and subsystem symmetry. As an application, we illustrate our results with several examples in one and two dimensions, which separate two different SPTs.
\end{abstract}
\maketitle

\section{Introduction}
One central task of condensed matter theory is to classify different phases and phase transitions.
In the past two decades, many exotic gapped phases beyond the Landau paradigm have attracted a considerable amount of interest.  A family of gapped quantum many-body systems which has been extensively studied is the SPT phase, such as the Haldane phase of the spin-1 antiferromagnetic Heisenberg model \cite{HALDANE1983464,PhysRevLett.59.799,PhysRevB.83.075103,PhysRevB.84.165139,PhysRevB.81.064439,PhysRevB.83.035107} and bosonic analogs
of topological insulators and superconductors \cite{chen2012symmetry,PhysRevX.3.011016}. In general, a nontrivial SPT phase can be characterized by a  short-range entangled ground state on a closed lattice \cite{PhysRevB.82.155138}, nonlocal string order \cite{PhysRevB.45.304,kennedy1992hidden,oshikawa1992hidden}, gapless edge  states \cite{PhysRevB.85.075125,PhysRevB.83.035107,PhysRevB.90.235137} and entanglement spectrum \cite{PhysRevB.81.064439,PhysRevB.83.035107}, which is protected by a global symmetry symmetry $G$. One systematical way to construct nontrivial SPT phases is via decorating domain wall \cite{Chen2013SymmetryprotectedTP,wang2021domain}, starting from the lower dimensional SPTs.  This domain wall decoration can be implemented
 through finite depth unitary operators \cite{yoshida2016topological}, which defines a duality relating different SPT phases by conjugating their Hamiltonians.

While the topological properties of the SPT phases are by
now largely well-understood, our understanding of phase transitions between them is still under development. Since an SPT phase does not break any symmetry, the phase transition between SPT phases is expected to host novel quantum critical behavior, which is beyond the Landau-Ginzburg-Wilson (LGW) paradigm. Recently, many analytic and numerical development displays deep connections between such phase transitions and deconfined quantum critical points (DQCP) \cite{PhysRevB.70.075104,PhysRevB.93.125101,PhysRevB.93.195141,PhysRevB.97.195115,senthil2019duality,PhysRevX.9.021034}, including the study of quantum criticality separating SPT phases in 1D and 2D \cite{PhysRevB.87.045129,PhysRevB.89.195143,PhysRevB.90.214426,PhysRevLett.115.237203,PhysRevB.97.125112,PhysRevB.96.165124,tantivasadakarn2021building,mudry2019quantum,aksoy2021stability}. However, this scheme
can not tell us dynamic properties of criticality from microscopic models, such as the deconfined degrees
of freedom. Thus an overarching theoretical framework of
such quantum criticality is still lacking. Moreover, the existing research mainly focuses on the phase transition between SPT phases which is protected by 0-form symmetry and classified by the group cohomology. However, the study of critical theory with generalized symmetry, such as higher form symmetry \cite{Kapustin2017,kapustin2014coupling,Gaiotto:2014kfa,Nussinov:2006iva,Batista:2004sc,Nussinov:2009zz,Nussinov:2011mz} and subsystem symmetry \cite{PhysRevB.98.035112,PhysRevB.98.235121,PhysRevResearch.2.012059,PhysRevResearch.4.L032008}, has remained relatively scarce.

In this work, we will focus on the system where the duality relating different SPT phases becomes a non-onsite symmetry, i.e., the system
is self-dual. The self-duality forces the system to stay
on the boundary separating duality-related
phases, often inducing critical or muticritical points. There have been several studies showing that this duality symmetry often shares an 't Hooft anomaly with the 0-form symmetry protecting SPT phases, based on the group cohomology fixed points wavefunctions and short-range entanglement properties of SPT states \cite{tsui2015quantum,PhysRevB.100.165132}.  Therefore the approach above strictly hold away from the critical point and
break down when the correlation length diverges. One can expect that this approach can still be applied to critical points and imply restrictions on the dynamical properties, since 't Hooft anomaly is preserved in the renormalization group (RG) flow  \cite{tHooft:1980xss}. Intuitively, this result can be understood as follows: the system with an 't Hooft anomaly is imposed with general constraints on its spectrum by the notion of ingappabilities \cite{lieb1961two,Affleck2004,PhysRevLett.78.1984,PhysRevLett.84.1535,PhysRevB.69.104431,PhysRevLett.118.021601,yao2019anomaly,aksoy2021lieb,li2022boundary}, namely the system cannot have a unique symmetric gapped ground state, which is consistent with the fact that duality operator connects different SPTs.

The work presented here is to provide another geometric approach which can be directly applied to any local self-dual Hamiltonian with additional onsite symmetry rather than basing on fixed-point Hamiltonians and wavefunctions. More precisely, we will prove the ingappability of the bulk spectrum of the self-dual model by making use of the spectrum robustness argument on the symmetry twisted boundary conditions (STBC) \cite{PhysRevB.98.155137,yao2021twisted,yao2022geometric}, which does not depend on the divergent behaviour of correlation length. Moreover, our  approach can also apply to the self-dual model with generalized symmetry, such as higher form and subsystem symmetry. Therefore, we can use this method to discuss the dynamical properties of phase transitions between SPT phases protected by generalized symmetry, which can provide insights to field theory and numerical study
to determine their properties.  In the main text, we shall restrict our attention to the system with several $\mathbb{Z}_2$ symmetry and prove the ingappabilities in a systematic manner. The generalization to $\mathbb{Z}_N$ symmetry is straightforward and the related discussion is provided in the appendix.

The organization of the paper is as follows.
	In section \ref{sec 1}, we discuss ingappabilities of the systems with duality and ($\mathbb{Z}_2)^{d+1}$  0-form symmetry in $d$ spatial dimensions. More precisely, we begin with the detail of proof for ingappabilities in one dimensional models and provide an intuitive argument for higher dimensional models, while the rigorous proof is left in the appendices. In section \ref{sec 2}, we apply a similar method to prove ingappabilities of two kinds of self-dual models. The first kind of model possesses the $\mathbb{Z}_2\times \mathbb{Z}_2$ subsystem symmetry which only acts on one dimensional sublattice, whose spatial dimension can be arbitrary. The other kind respects $\mathbb{Z}_2$ one-form symmetry and $\mathbb{Z}_2$ zero-form symmetry, which is defined on the two dimensional lattice.
	As an application of our framework, we present some concrete examples in one and two dimensions in section \ref{sec 3}. These models exhibit critical properties which are consistent with our general proof. In section \ref{sec 4}, we end with a conclusion and discussion of directions
for future studies.

\section{Ingappabilities of  duality and 0-form symmetry }\label{sec 1}

In this section, we will consider duality transformation which is used to construct bosonic SPT phases protected by a 0-form discrete symmetry. And we will study ingappabilities of the self-dual system where the duality becomes an emergent symmetry. More precisely, we focus on the $(\mathbb{Z}_2)^{d+1}$ symmetry on $d$-dimensional lattice and the duality operator $U_{(\underbrace{0,0,\cdots,0}_{d+1})}$ can be realized as multiqubit control-Z operators \cite{yoshida2016topological}. The discussion on general $\mathbb{Z}_N$ cases in one dimension will be provided in the appendix~\ref{SM1}.

\subsection{One dimensional model with  duality symmetry  and $\mathbb{Z}_{2}^A\times\mathbb{Z}_{2}^G$ onsite symmetry }\label{1dproof}

Let us warm up with a closed chain and assign two spin-$\frac{1}{2}$'s per unit cell: the spins $\sigma$ living on the integer sites are charged under $\mathbb{Z}_{2}^G$ while those living on the half-integer sites to be denoted $\tau$ are charged under $\mathbb{Z}_{2}^A$. The symmetry operators are defined to be
\begin{eqnarray}
U_A= \prod_{i=1}^{L} \tau_{i-\frac{1}{2}}^x, ~~~~~ U_G= \prod_{i=1}^{L} \sigma_{i}^x,
\end{eqnarray}
where $\sigma^a_{i}$ and $\tau^a_{i-\frac{1}{2}}$, $a=x,y,z$, are Pauli matrices acting on the two spin-$\frac{1}{2}$'s, and $L$ is the number of unit cells. The non-onsite $\mathbb{Z}_{2}^{(0,0)}$ duality symmetry is given by:
\begin{eqnarray}\label{UDW}
U_{(0,0)}=&&\prod^L_{i=1}\text{CZ}_{i-\frac{1}{2},i}\text{CZ}_{i,i+\frac{1}{2}}\nonumber\\=&&\prod_{i=1}^{L}\exp\left[\frac{\pi i}{4}(1-\sigma^{z}_{i})(1-\tau^{z}_{i-\frac{1}{2}})\right]\nonumber\\&&\prod_{i=1}^{L}\exp\left[\frac{\pi i}{4}(1-\sigma^{z}_{i})(1-\tau^{z}_{i+\frac{1}{2}})\right].
\end{eqnarray}

It is known that the $U_{(0,0)}$ transformation correspond to the  domain wall decoration \cite{scaffidi2017gapless,PhysRevB.97.165114,PhysRevLett.122.240605,Thorngren:2020wet,Chen2013SymmetryprotectedTP,Wang:2021nrp,li2022symmetry} with respect to  $\mathbb{Z}_2^G$ and  $\mathbb{Z}_2^A$ symmetry. For example we can identify the spin configuration representing the $\mathbb{Z}_2^A$ domain wall, i.e. $\tau_{i-\frac{1}{2}}^z \tau_{i+\frac{1}{2}}^z=-1$. Then on the link $(i-\frac{1}{2}, i+\frac{1}{2})$, the $U_{(0,0)}$ assigns the wavefunction an extra minus sign if $\sigma_{i}^z=-1$ on the wall.  Thus one assigns a minus sign to the wavefunction with two configurations $(\tau_{i-\frac{1}{2}}^z, \sigma_{i}^z, \tau_{i+\frac{1}{2}}^z)=(1,-1,-1), (-1,-1,1)$ and leaves it invariant for other configurations. Physically, this operation means that we stack a (0+1)d $\mathbb{Z}_2^G$ SPT on the link $(i-\frac{1}{2},i+\frac{1}{2})$ with nontrivial $\mathbb{Z}_2^A$ domain wall configuration. One can swap the place of $\sigma$ and $\tau$ spin in the explanation above, then $U_{(0,0)}$ operator must also decorate a nontrivial $\mathbb{Z}_2^G$ domain wall with a (0+1)d $\mathbb{Z}_2^A$ SPT at the same time.

To see how this duality connects different SPT phases, one can start with a trivial phase with a paramagnetic Hamiltonian:
\begin{eqnarray}\label{UDW}
H^{(0,0)}_{0}:=-\sum^L_{i=1}(\sigma^x_i+\tau^x_{i+\frac{1}{2}})
\end{eqnarray}
whose ground state is the product state:
\begin{eqnarray}
|\text{GS}\rangle=|\sigma^x_i=1, \tau^x_{i+\frac{1}{2}}=1\rangle.
\end{eqnarray}

Then the nontrivial SPT Hamiltonian is arrived at by conjugating the paramagnetic Hamiltonian by $U_{(0,0)}$,  yielding
\begin{eqnarray}\label{cluster}
H^{(0,0)}_1:= &&U_{(0,0)}H^{(0,0)}_{0}U_{(0,0)}^\dagger \nonumber\\=&& -\sum_{i=1}^{L} (\sigma^z_{i} \tau_{i+\frac{1}{2}}^x \sigma^z_{i+1} + \tau_{i-\frac{1}{2}}^z\sigma^x_{i} \tau_{i+\frac{1}{2}}^z)
\end{eqnarray}
which is known as the cluster model. This Hamiltonian also possesses a single ground state on a closed chain. However, if we consider the open boundary condition,  there are stable edge modes localized on the boundaries \cite{Chen2013SymmetryprotectedTP}.

Now, let us start to prove ingappabilities for any self-dual Hamiltonian, namely,
a one-dimensional spin-1/2 chain cannot have a unique gapped symmetric ground state if $\mathbb{Z}_2^{(0,0)}$ and $\mathbb{Z}^{A}_2 \times \mathbb{Z}^G_2 $ are strictly imposed. More precisely,  we consider a closed spin
chain of length $L$ with the periodic boundary condition
(PBC), which is used to eliminate possible edge modes, as we are only interested in the bulk spectra. However, instead of studying the spectra under PBC directly, let us introduce a Hamiltonian with the
$\mathbb{Z}_2^A$-symmetry twisted boundary condition (STBC):
\begin{eqnarray}\label{stbc}
\tau_{i+L-\frac{1}{2}}^{a}\equiv U_A \tau_{i-\frac{1}{2}}^{a}U^{-1}_A, \text{with} \ i=1,\cdots,L,
\end{eqnarray}
where the closed boundary bond is between the sites $i=L$ and $i=1$.

Similar to the reference~\cite{yao2021twisted}, this twisted Hamiltonian  explicitly breaks the original $\mathbb{Z}^{(0,0)}_2$ symmetry, but instead
is invariant only when followed by an additional ``gauge" transformation:
\begin{eqnarray}\label{1dduality}
U^{(1)}_{(0,0)}=\sigma^z_L U_{(0,0)}.
\end{eqnarray}

To see it, we begin with a local term $H_{j,\cdots,L,1/2,\cdots,k}$ in the Hamiltonian with PBC which crosses the boundary link $(L,1/2)$. Here the index $(j,\cdots,L,1/2,\cdots,k)$  means this term only acts on these sites at most, and the support of this term is bounded due to locality.
Thus after twisting, the resulting term when $k\in\mathbb{Z}+\frac{1}{2}$ is
\begin{eqnarray}
&&H^{\text{tw}}_{j,\cdots,L,1/2,\cdots,k}=(\prod^{k+\frac{1}{2}}_{i=1}\tau^x_{i-\frac{1}{2}})H_{j,\cdots,L,\frac{1}{2},\cdots,k}(\prod^{k+\frac{1}{2}}_{i=1}\tau^x_{i-\frac{1}{2}}),\nonumber\\~
\end{eqnarray}
while when $k\in\mathbb{Z}$, it is given by
\begin{eqnarray}
H^{\text{tw}}_{j,\cdots,L,1/2,\cdots,k}=(\prod^{k}_{i=1}\tau^x_{i-\frac{1}{2}})H_{j,\cdots,L,\frac{1}{2},\cdots,k}(\prod^{k}_{i=1}\tau^x_{i-\frac{1}{2}}).
\end{eqnarray}
The above equations do not imply that the entire Hamiltonians $H^\text{tw}$ and $H$ are related by a unitary transformation. 
If they were related by some unitary transformation,
then their spectra would be identical, which is obviously incorrect.
Formally,
they are related by an ``ill-defined'' transformation $\prod_{i=1}^\infty\tau_{i-1/2}^x$. However, 
as manifested in the above equations,
the long tail until the formal ``$\infty$'' is invisible to any local term in the Hamiltonian since the interaction range of the local terms is bounded by a fixed finite number.
Due to this locality condition,
the action of this ill-defined transformation on the Hamiltonian terms becomes well-defined.

Then we can consider the dual term of this local term, and under the original periodic boundary condition, it is invariant up till a change of subscript:
\begin{eqnarray}
{H}_{j-\frac{1}{2},\cdots,L,1/2,\cdots,k+\frac{1}{2}}=U_{(0,0)}H_{j,\cdots,L,1/2,\cdots,k}U_{(0,0)}^{\dagger}.
\end{eqnarray}
We notice that this term acts on the region $(j-1/2,j,\cdots,L,1/2,\cdots,k,k+1/2)$ at most as $U_{(0,0)}$ only acts on two nearest neighboured sites. 
Imposing the modified transformation onto the twisted local term, 
the resulting term when $k\in\mathbb{Z}+\frac{1}{2}$ is
\begin{eqnarray}\label{eq:1d twist-1}
&&U^{(1)}_{(0,0)}{H}^{\text{tw}}_{j,\cdots,L,1/2,\cdots,k}({U^{(1)}_{(0,0)}})^\dagger\nonumber\\&=&\sigma_L^zU_{(0,0)}(\prod^{k+\frac{1}{2}}_{i=1}\tau^x_{i-\frac{1}{2}})H_{j,\cdots,L,\frac{1}{2},\cdots,k}(\prod^{k+\frac{1}{2}}_{i=1}\tau^x_{i-\frac{1}{2}})U^{\dagger}_{(0,0)}\sigma_L^z\nonumber\\
&=&(\prod^{k+\frac{1}{2}}_{i=1}\tau^x_{i-\frac{1}{2}})U_{(0,0)}H_{j,\cdots,L,\frac{1}{2},\cdots,k}U^{\dagger}_{(0,0)}(\prod^{k+\frac{1}{2}}_{i=1}\tau^x_{i-\frac{1}{2}})\nonumber\\
&=&H^{\text{tw}}_{j-\frac{1}{2},\cdots,L,1/2,\cdots,k+\frac{1}{2}}.
\end{eqnarray}
When $k\in\mathbb{Z}$, it is given by
\begin{eqnarray}\label{eq:1d twist-2}
&&U^{(1)}_{(0,0)}H^{\text{tw}}_{j,\cdots,L,1/2,\cdots,k}(U^{(1)}_{(0,0)})^{\dagger}\nonumber\\
&&=\sigma^z_L U_{(0,0)}(\prod^{k}_{i=1}\tau^x_{i-\frac{1}{2}})H_{j,\cdots,L,\frac{1}{2},\cdots,k}(\prod^{k}_{i=1}\tau^x_{i-\frac{1}{2}})U^{\dagger}_{(0,0)}\sigma^z_L\nonumber\\
&&=(\prod^{k}_{i=1}\tau^x_{i-\frac{1}{2}})U_{(0,0)}H_{j,\cdots,L,\frac{1}{2},\cdots,k}U_{(0,0)}(\prod^{k}_{i=1}\tau^x_{i-\frac{1}{2}})\nonumber\\&&=H^{\text{tw}}_{j-\frac{1}{2},\cdots,L,1/2,\cdots,k+\frac{1}{2}}.
\end{eqnarray}
Let us give a brief explanation of Eq.~\eqref{eq:1d twist-1} and Eq.~\eqref{eq:1d twist-2}.  When we exchange $U_{(0,0)}$ and the $\tau^x$ string operator, the $\sigma^z$ operators on two endpoints of  this string are generated. Since the Hamiltonian term is local,  we can consider a long enough string to twist terms that cross the boundary link. The $\sigma^z$ at the rightmost site does not touch these local terms and only $\sigma^z_{L}$  appears in the modified $\mathbb{Z}^{(0,0)}_2$ symmetry.

In the next step, we consider the local term $H_{j,\cdots,L}$ which ends at the boundary. Thus it is unchanged after twisting. Moreover, the dual term is as follows
\begin{eqnarray}
H_{j-\frac{1}{2},\cdots,L,\frac{1}{2}}=U_{(0,0)}H_{j,\cdots,L}U^{\dagger}_{(0,0)}.
\end{eqnarray}
After twisting, we can obtain
\begin{eqnarray}
U^{(1)}_{(0,0)}H^{\text{tw}}_{j,\cdots,L}(U^{(1)}_{(0,0)})^{\dagger}&&=U_{(0,0)}\sigma^z_{L} H_{j,\cdots,L}\sigma^z_{L} U^{\dagger}_{(0,0)}\nonumber\\&&=\tau^x_{\frac{1}{2}}U_{(0,0)}H_{j,\cdots,L}U^{\dagger}_{(0,0)}\tau^x_{\frac{1}{2}}\nonumber\\
&&=H^{\text{tw}}_{j-\frac{1}{2},\cdots,L,\frac{1}{2}}.
\end{eqnarray}

As last, we consider the local term $H_{j,\cdots,k}$ $(j<k<L)$ and its  dual term $U_{(0,0)}H_{j,\cdots,k}U^{\dagger}_{(0,0)}$. Since both of them do not cross the boundary, they are unchanged after twisting. Moreover, we also notice that  $U_{(0,0)}H_{j,\cdots,k}U^{\dagger}_{(0,0)}=U^{(1)}_{(0,0)}H_{j,\cdots,k}(U^{(1)}_{(0,0)})^{\dagger}$. Thus the local term above and its dual term after twisting are also related by $U^{(1)}_{(0,0)}$ operator. Thus we conclude that the twisted Hamiltonian is invariant under this modified duality operator which finishes proof of Eq.\eqref{1dduality}.

Since $U^{(1)}_{(0,0)}$ anticommucates with the $U_G$ operator, the twisted Hamiltonian must have an exactly double degenerate spectrum. Then we arrive at a rigorous conclusion:
If a 1d $\mathbb{Z}^{A}_2 \times \mathbb{Z}^G_2$-invariant spin-1/2 chain also possesses $\mathbb{Z}_2^{(0,0)}$ symmetry, it must have a doubly degenerate spectrum under STBC \eqref{stbc}.

However, our initial interest is the low energy spectrum under PBC. Fortunately, there is a spectrum robustness theorem relating STBC and PBC \cite{PhysRevB.98.155137,yao2021twisted,yao2022geometric}: if the (pre-twisted) Hamiltonian under PBC has a unique
and gapped ground state, the twisted Hamiltonian under
STBC also possesses a unique gapped ground state. As a direct consequence, we can obtain that any Hamiltonian under PBC must either be gapless or have a nontrivial ground-state
degeneracy, rather than a unique gapped ground state,
which finishes the proof of the ingappability of bulk spectra.

Here we remark that if the system under PBC is in the SSB phase, then either the global onsite symmetry is broken or the duality symmetry is broken. The first case will be discussed in section~\ref{sec 3}, where SSB phases can be detected by the expectation value and correlation function of local order parameters. Besides, if only the duality symmetry is broken, the degenerate ground states should have the same topological response as, respectively, the trivial and the nontrivial SPT states, 
i.e., it can be detected by the string order parameter.
Nevertheless, a concrete example of such an SSB can be a future interest.

\subsection{Two dimensional model with  duality symmetry  and $(\mathbb{Z}_{2})^3$ onsite symmetry }\label{2dproof}
\begin{figure}[htpb] 
\centering 
\includegraphics[width=0.35\textwidth]{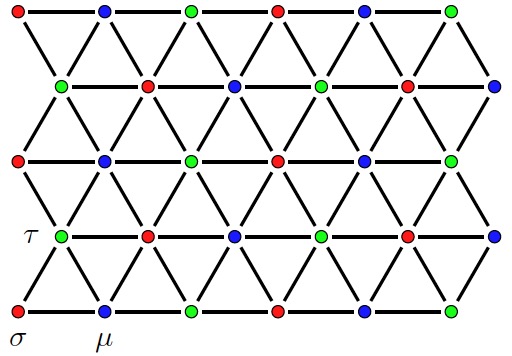} 
\caption{ The triangular lattice where spins on the vertices are colored in red, green and blue. The  symmetry is generated by spin flips on each sublattice. } 
\label{Triangle} 
\end{figure}

In 2+1 dimensions, we focus on the triangle lattice shown in Fig.~\ref{Triangle} and assign a spin $\frac{1}{2}$ on each vertex.

Then one can naturally consider the onsite $(\mathbb{Z}_2)^3$ symmetry generated by spin-flips on each of the three sublattices, which are colored by red, green and blue:
\begin{eqnarray}
U_A= \prod_{v\in A} \sigma_v^x, ~~~~~ U_B= \prod_{v\in B} \tau_{v}^x, ~~~~~ U_C= \prod_{v\in C} \mu_{v}^x.
\end{eqnarray}
Here we label sublattice by $A$, $B$ and $C$.

The duality operator connecting trivial and nontrivial SPT phases protected by this $(\mathbb{Z}_2)^3$ symmetry is given by~\cite{yoshida2016topological}
\begin{eqnarray}
&&U_{(0,0,0)}=\prod_{(i,j,k)\in\bigtriangleup}(\text{CCZ})_{i,j,k},\nonumber\\ &&(\text{CCZ})_{i,j,k}|\alpha,\beta,\gamma\rangle_{i,j,k}=(-1)^{\alpha\beta\gamma}|\alpha,\beta,\gamma\rangle_{i,j,k},
\end{eqnarray}
where $\bigtriangleup$ represents the sets of all  triangles and CCZ is a unitary operator acting on each triple of spins belonging to one triangle. Moreover, $\alpha$, $\beta$ and $\gamma$ belong to $\{0,1\}$ and they represent the spins of $i,j,k$ sites where spin up corresponds to 0 and spin down corresponds to 1.

\begin{figure}[htpb] 
\centering 
\includegraphics[width=0.25\textwidth]{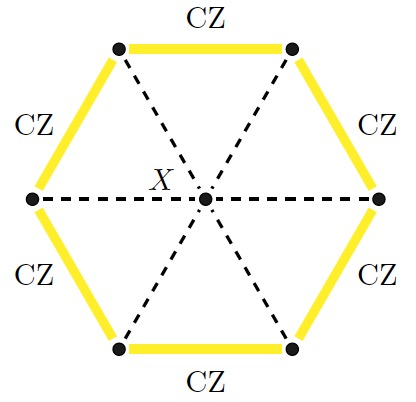} 
\captionof{figure}{The local
		term $O_v$ where a Pauli $X$ operator at the vertex $v$ is decorated by CZ operators on all links surround this vertex. }\label{localterm}
\end{figure}

To see the effect of the duality operator, one can also start with the paramagnetic  Hamiltonian which is in a trivial SPT phase:
\begin{eqnarray}
H^{(0,0,0)}_{0}=-\sum_{v\in A}\sigma^x_v-\sum_{v\in B}\tau^x_v-\sum_{v\in C}\mu^x_v.
\end{eqnarray}
Then the nontrivial SPT Hamiltonian is obtained by conjugating the paramagnetic Hamiltonian by $U_{(0,0,0)}$,  yielding
\begin{eqnarray}\label{CCZ Hamiltonian}
&&H^{(0,0,0)}_1:= U_{(0,0,0)}H^{(0,0,0)}_{0}U^{\dagger}_{(0,0,0)} = -\sum_v O_v, \\&& O_{v\in A}=\sigma^x_v \prod_{e\in 1-\text{link}(v)}\text{CZ}_e,\nonumber\\&& O_{v\in B}=\tau^x_v \prod_{e\in 1-\text{link}(v)}\text{CZ}_e,\nonumber\\&& O_{v\in C}=\mu^x_v \prod_{e\in 1-\text{link}(v)}\text{CZ}_e.
\end{eqnarray}
Here $O_v$ operator is represented in Fig.~\ref{localterm} including CZ operators on all links $e$ surround  the vertex $v$. These links are also known as the 1-links of vertex $v$ \cite{yoshida2016topological}. Moreover, $X$ operator corresponds to the Pauli operator $\sigma^x$ or $\tau^x$ or $\mu^x$.


In fact, there are seven distinct generators for SPT phases protected by $\mathbb{Z}_2\times\mathbb{Z}_2\times\mathbb{Z}_2 $ symmetry according to the group cohomology, which
can be classified into three types, named type-I, type-II and type-III. The Hamiltonian \eqref{CCZ Hamiltonian} in our interest corresponds to the type-III  which generates the class $H^1\left(\mathbb{Z}_2,H^2(\mathbb{Z}_2\times \mathbb{Z}_2, U(1))\right)$.  Physically, it can be understood as decorating a one dimension SPT phase protected by the last two $\mathbb{Z}_2$ symmetries on the domain wall of the first
$\mathbb{Z}_2$  symmetry \cite{Chen2013SymmetryprotectedTP}.

\begin{figure}[htpb] 
\centering 
\includegraphics[width=0.35\textwidth]{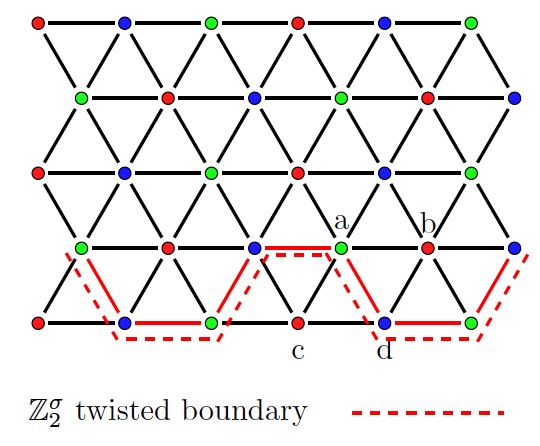} 
\captionof{figure}{ Twisted boundary condition by $\mathbb{Z}^{\sigma}_2$ symmetry. The red solid links represent CZ operators of $\tau$ and $\mu$ spins in  $U^{(1)}_{(0,0,0)}$. }\label{sigmatwist}
\end{figure}

Now, let us discuss the ingappability of the bulk spectrum of the self-dual model. Similar to the section \ref{1dproof}, we first twist the boundary condition by $\mathbb{Z}^{\sigma}_2$ symmetry and the closed boundary is an armchair line shown in Fig.~\ref{sigmatwist}.  Then the twisted Hamiltonian explicitly breaks $U_{(0,0,0)}$ but possesses a modified symmetry 
\begin{eqnarray}\label{duality2d}
U^{(1)}_{(0,0,0)}=U_{(0,0,0)}U_{(0,0)}(S_{\text{red}}),
\end{eqnarray}
where $S_{\text{red}}$ represents the red solid line in Fig.~\ref{sigmatwist}.

\begin{figure}[htpb] 
\centering 
\includegraphics[width=0.37\textwidth]{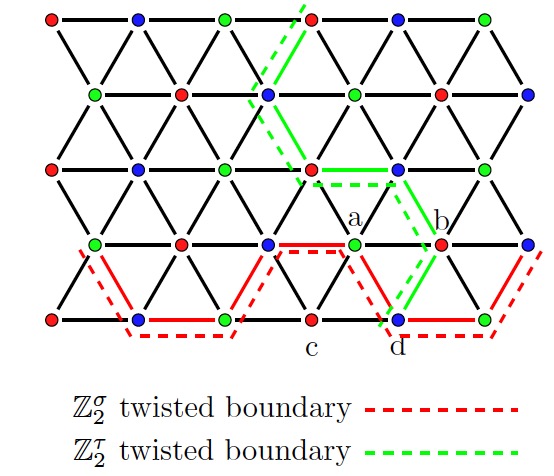} 
\captionof{figure}{  Twisted boundary condition by $\mathbb{Z}^{\sigma}_2$ symmetry and $\mathbb{Z}^{\tau}_2$ symmetry . The green solid links represent CZ operators of $\sigma$ and $\mu$ spins in the $U^{(2)}_{(0,0,0)}$. }\label{tautwist}
\end{figure}

In the next step, we can continue to consider the STBC in another direction where the addition closed boundary is the green armchair $\mathbb{Z}^{\tau}_{2}$ line in Fig.~\ref{tautwist}.

After twisting, both two operators in $U^{(1)}_{(0,0)}$ will be modified according to Eq.\eqref{1dduality} and Eq.\eqref{duality2d}. Thus the modified duality transformation is as follows
\begin{eqnarray}
U^{(2)}_{(0,0,0)}=U_{(0,0,0)}U_{(0,0)}(S_{\text{green}})U_{(0,0)}(S_{\text{red}})\mu^z_{d}.
\end{eqnarray}
Here $S_{\text{green}}$ represents the green solid line and $\mu^z_{d}$ comes from modifying of $U_{(0,0)}(S_{\text{red}})$.

Since $U^{(2)}_{(0,0,0)}$ anticommutes with $U_C$, all the eigenstates of the STBC Hamiltonian are doubly degenerate. Then according to the robustness of the spectrum under STBC, this result implies an ingappability of
the system under PBC, i.e., a unique gapped ground state is forbidden. Hence, the only task is to prove Eq.\eqref{duality2d}. We will provide an intuitive proof on an infinite lattice without boundaries, which should be valid for a closed lattice in the thermodynamic limit. A rigorous proof for the periodic boundary condition will be provided in the appendix~\ref{SM2}.

We first span the ground state by the eigenstate of $Z$ operators:
\begin{eqnarray}
|\text{GS}\rangle=\sum_{(i,j,k)\in\bigtriangleup/\bigtriangledown}\psi_{ijk}|\alpha_1,\alpha_2,\alpha_3\rangle_{ijk}.
\end{eqnarray}

Since the lattice now is not closed, the twisted  "boundary condition" in the Fig.~\ref{sigmatwist} is equivalent to adding a same armchair $\mathbb{Z}^{\sigma}_2$ twisting line. More precisely, the twisted Hamiltonian  and original Hamiltonian are related by a unitary transformation: $H^{\mathbb{Z}^{\sigma}_2}_{\text{tw}}=(\prod_{v\in K}\sigma^x_v)H (\prod_{v\in K}\sigma^x_v)$ where the $K$ is the set of all $A$ sites below this twisting line. And the corresponding ground state is $|\text{GS}\rangle^{\mathbb{Z}^{\sigma}_2}_{\text{tw}}=\prod_{v\in K}\sigma^x_v|\text{GS}\rangle$ . Moreover, the modified duality operator becomes $U^{(1)}_{(0,0,0)}=(\prod_{v\in K}\sigma^x_v)U_{(0,0,0)}(\prod_{v\in K}\sigma^x_v)$.

Let us focus on the action of the duality operator on the two neighboring triangles which consist of $a$,$b$,$c$ and $d$ sites in Fig.~\ref{sigmatwist}. The phase of the original duality operator is given by  
\begin{eqnarray}
&&U_{(0,0,0)}|\alpha_1,\alpha_2,\alpha_3,\alpha_4\rangle\nonumber\\=&&(-1)^{(\alpha_2+\alpha_3)\alpha_1 \alpha_4}|\alpha_1,\alpha_2,\alpha_3,\alpha_4\rangle,
\end{eqnarray}
where $\alpha_1$, $\alpha_2$, $\alpha_3$ and $\alpha_4$ represents the spins of $a,b,c$ and $d$ sites.

After inserting this red twisting line, we need to flip the spin of $c$ site in these two triangles. Then local action of the duality operator is:
\begin{eqnarray}
&&U^{(1)}_{(0,0,0)}|\alpha_1,\alpha_2,\alpha_3,\alpha_4\rangle\nonumber\\=&&\sigma^x_c U_{(0,0,0)}\sigma^x_c|\alpha_1,\alpha_2,\alpha_3,\alpha_4\rangle\nonumber\\=&&U_{0,0,0}\text{CZ}_{a,d}|\alpha_1,\alpha_2,\alpha_3,\alpha_4\rangle,
\end{eqnarray}
where $\text{CZ}_{a,d}$ acts on the solid red link as a domain wall of $b$ and $c$ sites. We can sum over all pairs of neighboured triangles which are 
separated by the twisting line and the total addition phase is $U_{(0,0)}$ on solid red links in Fig.~\ref{sigmatwist}. This region is also an armchair line and is locally parallel to the twisting line. Therefore, the modified duality operator after adding the $\mathbb{Z}^{\sigma}_2$ symmetry twisting line is Eq.\eqref{duality2d}, which finishes our proof.

\subsection{Higher dimensional model with duality symmetry  and onsite symmetry}\label{anydproof}
Finally, we can  generalize our discussion to the ingappability for $(d+1)$-dimensional model with $(\mathbb{Z}_2)^{d+1} $ onsite symmetry and  duality symmetry. Let us consider a $d$-dimensional simplicial lattice which is $(d+ 1)$-colorable with color labels $a_1, \cdots, a_{d+1}$ and place a spin-$\frac{1}{2}$ on each vertex. One can naturally define a $(\mathbb{Z}_2)^{d+1} $ onsite symmetry associated to spin-flips on each sublattice $a_i$

\begin{eqnarray}
U_{a_i}=\prod_{\nu}X^{(a_i)}_\nu.
\end{eqnarray}

Then the duality operator is defined as
\begin{eqnarray}
&&U_{(\underbrace{0,0,\cdots,0}_{d+1})}=\prod_{(i_1, i_2,\cdots i_{d+1})\in\bigtriangleup}(\text{C}^{\otimes d}\text{Z})_{i_1, i_2,\cdots i_{d+1}},\nonumber\\ &&(\text{C}^{\otimes d}\text{Z})_{i_1, i_2,\cdots i_{d+1}}|\alpha_1,\alpha_2,\cdots,\alpha_{d+1}\rangle_{i_1, i_2,\cdots i_{d+1}}\nonumber\\=&&(-1)^{\prod^{d+1}_{j=1}\alpha_{j}}|\alpha_1,\alpha_2,\cdots,\alpha_{d+1}\rangle_{i_1, i_2,\cdots i_{d+1}},
\end{eqnarray}
where $\bigtriangleup$ represents the sets of all $d$-simplexes and $\alpha_k=0/1$ represents the spin up/down on the site $k$. 

Similar to the argument in the section \ref{2dproof}, we begin with the boundary condition twisted by the $U_{a_1}$ operator. More precisely, we consider a closed and connected $d-1$ dimensional sublattice $S_1$ which consists of $d-1$ dimensional simplex colored in $a_2,\cdots,a_{d+1}$. The $d-1$ dimensional twisted boundary is placed close and locally parallel to the $d-1$ dimensional sublattice above. Then the twisted Hamiltonian possesses a modified duality symmetry 
\begin{eqnarray}\label{dualityanyd}
U^{(1)}_{(\underbrace{0,0,\cdots,0}_{d+1})}=U_{(\underbrace{0,0,\cdots,0}_{d+1})}U_{(\underbrace{0,0,\cdots,0}_{d})}(S_1).
\end{eqnarray}

Next, we continue to twist the boundary condition by the $U_{a_2}$ operator similarly and the twisted boundary is close and locally parallel to  a closed and connected $d-1$ dimensional sublattice $S_2$ which consists of $d-1$ dimensional simplex colored in $a_1,a_3,\cdots,a_{d+1}$. According to Eq.\eqref{dualityanyd}, this twisted Hamiltonian possesses a new modified duality symmetry 
\begin{eqnarray}
&&U^{(2)}_{(\underbrace{0,0,\cdots,0}_{d+1})}\nonumber\\=&&U_{(\underbrace{0,0,\cdots,0}_{d+1})}U_{(\underbrace{0,0,\cdots,0}_{d})}(S_2)U_{(\underbrace{0,0,\cdots,0}_{d})}(S_1)\nonumber\\ &&U_{(\underbrace{0,0,\cdots,0}_{d-1})}(S_1 \cap S_2).
\end{eqnarray}
Here $S_1 \cap S_2$ is a closed and connected $d-2$ dimensional sublattice $S_2$ which consists of $d-1$ dimensional simplex colored in $a_3,\cdots,a_{d+1}$.

Moreover, we can continue to twist the boundary condition by the $U_{a_3}, U_{a_4},\cdots, U_{a_d}$ operators step by step.   The final twisted Hamiltonian possesses a new modified duality symmetry
\begin{eqnarray}\label{finalduality}
&&U^{(d)}_{(\underbrace{0,0,\cdots,0}_{d+1})}\nonumber\\=&&U_{(\underbrace{0,0,\cdots,0}_{d+1})}\prod^{d}_{k=1}\prod_{\{i_{\alpha}\}}U_{(\underbrace{0,0,\cdots,0}_{d+1-k})}(\cap^k_{\alpha=1} S_{i_{\alpha}}).\nonumber\\~
\end{eqnarray}

When $k<d$, the operator $U_{(\underbrace{0,0,\cdots,0}_{d+1-k})}(\cap^k_{\alpha=1} S_{i_{\alpha}})$ commutes with all onsite symmetries since the sublattice $\cap^k_{\alpha=1} S_{i_{\alpha}}$ is closed. However when $k=d$, this operator is the product of $Z$ operators in the sublattice  $ \cap^d_{i=1} S_i$. We can assume $ \cap^d_{i=1} S_i$ is one site colored in $a_{d+1}$  and thus the final modified duality operator \eqref{finalduality} anticommutes with $U_{d+1}$. Then all the eigenstates of the final twisted
Hamiltonian are doubly degenerate which implies an ingappability of
the spectra under PBC.

    To prove Eq.\eqref{dualityanyd}, we also provide an intuitive proof on the infinite lattice without boundaries here and the rigorous proof for the periodic boundary condition is left in the appendix~\ref{SM2}. Now, the twisted "boundary condition" is equivalent to adding a same armchair twisting line. Similarly, we only need to focus on the action of  the duality operator on the two neighboured $d$ dimensional simplex
separated by the twisting line. We denote that one simplex has sites $(i_1, i_2,\cdots i_{d+1})$ and the other has sites  $(i_2, i_3,\cdots i_{d+2})$. Thus they share a $d-1$ dimensional simplex with sites $(i_2, i_3,\cdots i_{d+1})$ which belongs to $S_1$. Since the sites of each $d$ dimensional simplex are in different colors, the sites $(i_2, i_3,\cdots i_{d+1})$ has different colors and $i_1$ and $i_{d+2}$ share the same color.    The phase of the duality operator is given by  
\begin{eqnarray}
&&U_{(\underbrace{0,0,\cdots,0}_{d+1})}|\alpha_1,\alpha_2,\cdots,\alpha_{d+2}\rangle\nonumber\\=&&(-1)^{(\alpha_1+\alpha_{d+2})\prod^{d+1}_{j=2}\alpha_j}|\alpha_1,\alpha_2,\cdots,\alpha_{d+2}\rangle.
\end{eqnarray}
Then if we insert a twisting line between the sites $i_1$ and $i_{d+2}$, the duality operator is  twisted locally as:
\begin{eqnarray}
&&U^{(1)}_{(\underbrace{0,0,\cdots,0}_{d+1})}|\alpha_1,\alpha_2,\cdots,\alpha_{d+2}\rangle\nonumber\\=&&X^{(a_1)}_{i_1}U_{(\underbrace{0,0,\cdots,0}_{d+1})}X^{(a_1)}_{i_1}|\alpha_1,\alpha_2,\cdots,\alpha_{d+2}\rangle\nonumber\\=&&U_{(\underbrace{0,0,\cdots,0}_{d+1})}(\text{C}^{\otimes d-1}\text{Z})_{i_2, i_3,\cdots i_{d+1}}|\alpha_1,\alpha_2,\cdots,\alpha_{d+2}\rangle.\nonumber\\
\end{eqnarray}
  We can sum over all $\text{C}^{\otimes d-1}\text{Z}$ on the $d-1$ dimensional simplex for all pairs of neighboured $d$ dimensional simplex separated by the twisting line and the total addition phase is $U_{(\underbrace{0,0,\cdots,0}_{d})}(S_1)$.
 Therefore, the modified duality operator after symmetry twisting is Eq.\eqref{dualityanyd}.
\section{Ingappabilities of duality and higher form or subsystem symmetry}\label{sec 2}
In this section, we will discuss the ingappability of the bulk spectra of lattice models with higher form or subsystem symmetry and duality symmetry. Similarly, this duality operator also connects trivial and nontrivial SPT phases.  We will focus on the $\mathbb{Z}_2$ case, but the generalization to $\mathbb{Z}_N$ case should be straightforward.
\subsection{Ingappabilities of duality symmetry and $\mathbb{Z}_2 \times \mathbb{Z}_2$ line symmetry}
 We begin with a generalization of the result in section \ref{1dproof} to spin system invariant under duality and $\mathbb{Z}_2 \times \mathbb{Z}_2$ line symmetry which is a special subsystem symmetry. 

 Let us consider a $d$ dimensional body-centered cubic (BBC) lattice. The BCC lattice can be regarded as two displaced simple cubic lattices and we denote them as A/B sublattice. They live in the center of cubes of each other. Hence we can naturally assign two kinds of spin-$\frac{1}{2}$'s: the spin-$\frac{1}{2}$'s $\sigma$ living on the A sublattice are charged under $\mathbb{Z}_{2}^G$ while those living on the B sublattice are charged under $\mathbb{Z}_{2}^A$ and denoted as $\tau$. The $\mathbb{Z}_2 \times \mathbb{Z}_2$ line symmetry is generated by spin flips of each straight line and the number of generators is proportional to $L^{d-1}$. Furthermore, the non-onsite decorated domain wall duality operator is given by:
\begin{eqnarray}
U_{DW}=&&\prod_{<i,j>}\exp\left[\frac{\pi i}{4}(1-\sigma^{z}_{i})(1-\tau^{z}_{j})\right]
\end{eqnarray}
where $<i,j>$ is a pair of nearest neighboured sites.

 As before, one can start with the trivial cubic paramagnetic Hamiltonian:
\begin{eqnarray}\label{trivialsspt}
H_{\text{cubic}}:=-\sum_{i\in A}\sigma^x_i-\sum_{i\in B}\tau^x_i.
\end{eqnarray}
After conjugating this Hamiltonian by duality operator, one can obtain nontrivial subsystem symmetry protected topological (SSPT) Hamiltonian: 
\begin{eqnarray}
H_{\text{SSPT}}=&& U_{DW}H_{\text{cubic}}U_{DW}^\dagger \nonumber\\=&& -\sum_{C_A,j\in C_A} \tau^x_{j}\prod_{i\in C_A}\sigma^z_{i}  -\sum_{C_B,j\in C_B} \sigma^x_{j}\prod_{i\in C_B}\tau^z_{i} .\nonumber\\~
\end{eqnarray}
Here $C_A (C_B)$ refers to the cube of $A (B)$ sublattice including its vertex and center.

\begin{figure}[htpb] 
\centering 
\includegraphics[width=0.45\textwidth]{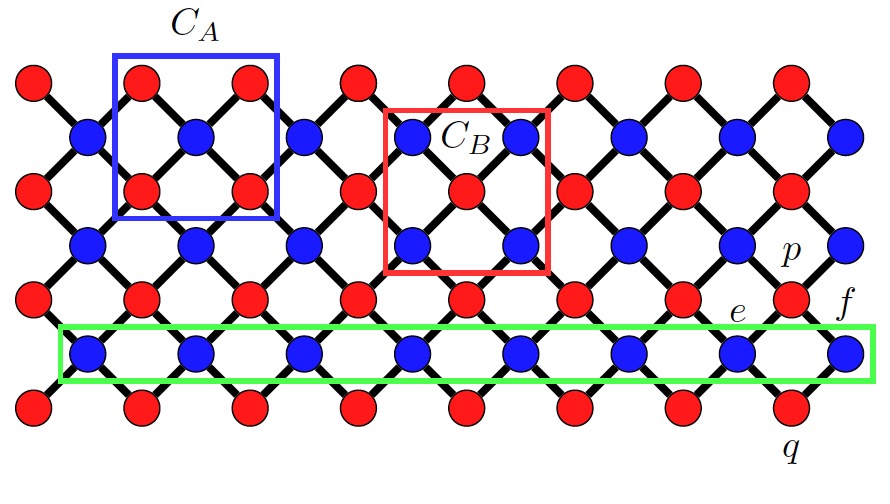} 
\captionof{figure}{The local terms in the SSPT Hamiltonian. $\tau$ and $\sigma$ spins live in blue and red sites respectively.}\label{BBC}
\end{figure}

In fact, when $d=1$, the subsystem symmetry is the ordinary $\mathbb{Z}^A_2\times\mathbb{Z}^G_2$ symmetry and the duality operator and SPT Hamiltonian are the same as the section \ref{1dproof}. If we take one step further and consider $d=2$, the BBC lattice is shown in Fig.~\ref{BBC} and the Hamiltonian of nontrivial SSPT phase is \cite{PhysRevB.98.035112}:
\begin{eqnarray}\label{2dsspt}
H^{2d}_{\text{SSPT}} = -\sum_{iklm\in C_A} \tau^x_{i}\sigma^z_{j}\sigma^z_{k}\sigma^z_{l}\sigma^z_{m} -\sum_{ijklm\in C_B} \sigma^x_{i}\tau^z_{j}\tau^z_{k}\tau^z_{l}\tau^z_{m} .\nonumber\\~
\end{eqnarray}
The first term 
involves the four $\sigma^z$ spins and the $\tau^x$
in the blue plaquette. The second term involves four $\tau^z$
spins and the $\sigma^x$ in the red plaquette. 

Moreover, if we combine the Hamiltonians \eqref{2dsspt} and \eqref{trivialsspt} in two dimensions, it has been shown that this self-dual model is in a first order phase transition  separating the nontrivial SSPT phase and
trivial paramagnetic phase \cite{PhysRevLett.93.047003,PhysRevLett.102.077203,PhysRevA.86.022317,Zhou:2022eig}. 

Now let us start to prove the ingappability of the general self-dual model where $U_{DW}$ becomes a symmetry and  $\mathbb{Z}_2 \times \mathbb{Z}_2$ line symmetry is imposed. Similar to the section above, we choose one straight line and twist the boundary condition by the spin flip operator supported by it.

For simplicity, we assume $\tau$ spin's coordinates are all integer and $\sigma$ spin's coordinates are all half integer. Without loss of generality, we choose a horizontal line of $\tau$ spins whose coordinates except $x_1$ are all zero.   Moreover, the twisted link is put between sites $x_1=L$ and $x_1=1$.   Then following the same calculation in the section \ref{1dproof}, the  twisted Hamilton possesses a modified duality symmetry by a "gauge" transformation: 
\begin{eqnarray}\label{lineduality}
U^{(1)}_{DW}=U_{DW}\prod^{2^{d-2}}_{i=1}\sigma^z_{i},
\end{eqnarray}
where the coordinate of site $i$ is  $(x_1=L-\frac{1}{2}, \pm \frac{1}{2},\cdots, \pm \frac{1}{2})$.

For example, when $d=2$, we can choose the straight line $l$ in the green frame and twist the boundary condition by the operator $\prod_{i\in l}\tau^x_i$. The twisted link can be put between site $e$ and $f$.  Then one can directly show the modified $U_{DW}$ is 
\begin{eqnarray}
U^{(1)}_{DW}= U_{DW}\sigma^z_p\sigma^z_q.
\end{eqnarray}

It is obvious that the modified duality operator Eq.\eqref{lineduality} anticommutes with $\mathbb{Z}^{\sigma}_2$ spin flip supported by the horizontal  line whose coordinates except $x_1$ are $\pm \frac{1}{2}$ . Thus we conclude that
if a  $\mathbb{Z}_2 \times \mathbb{Z}_2$ line symmetry-invariant spin-1/2 model also possesses $\mathbb{Z}_2^{DW}$ duality symmetry, it must have a doubly degenerate spectrum under STBC.
Then according to the spectrum robustness theorem connecting PBC and STBC,  we can arrive at ingappabilities of the bulk spectra of Hamiltonian under PBC. Moreover, we remark that this requirement for the ingappability can be relaxed where the lattice model only needs to preserve the $\mathbb{Z}_2 \times \mathbb{Z}_2$ subsystem symmetry supported on two nearest neighboured parallel lines and $\mathbb{Z}^{DW}_2$ duality symmetry.

\subsection{Ingappabilities of duality symmetry and $\mathbb{Z}_2$ 0-form and  $\mathbb{Z}_2$ 1-form symmetry}
Besides the subsystem symmetry, we can also discuss ingappabilities of the two dimensional lattice model which preserves a $\mathbb{Z}_2$ 0-form and  $\mathbb{Z}_2$ 1-form symmetry and duality symmetry. 

Let us consider a two dimensional body-centered square lattice shown in Fig.~\ref{(0,1)form}. But now we place $\tau$ spins colored in blue on each link whose coordinates are integer and half-integer and $\sigma$ spins colored in red on the center of each plaquette whose coordinates are both half-integers. 

The 0-form $\mathbb{Z}_2$ symmetry operator is generated by the spin flip of $\sigma$:
\begin{eqnarray}
U_G= \prod^L_{i,j=1} \sigma_{i+\frac{1}{2},j+\frac{1}{2}}^x.
\end{eqnarray}
The 1-form symmetry operator is a string of $\tau^x$ operators supported on a non-contractible loop $M$:
\begin{eqnarray}
U_A(M)= \prod_{e\cap M\ne\varnothing} \tau_{e}^x.
\end{eqnarray}
Here $e\cap M\ne\varnothing$ means the link $e$ intersects with the loop $M$.

Moreover, two configurations $|\{\tau^z_{e}\}\rangle$ and  $|\{\tilde{\tau}^z_{e}\}\rangle$ are gauge equivalent if they are related by a $\mathbb{Z}_2$ gauge transformation:
\begin{eqnarray}
\tilde{\tau}^z_{e}=W_m\tau^z_{e}W^{-1}_n,
\end{eqnarray}
where $m,n$ are sites on the boundary of link $e$ and $W_{m/n}=\pm 1$.
Thus, the elements of $\mathbb{Z}_2$ 1-form symmetry group  on the torus correspond to the cohomology group $H^1(\mathbb{T}^2,\mathbb{Z}_2)$ after  
considering the $\mathbb{Z}_2$ gauge
equivalence class, which differs from the subsystem symmetry.

\begin{figure}[htpb] 
\centering 
\includegraphics[width=0.4\textwidth]{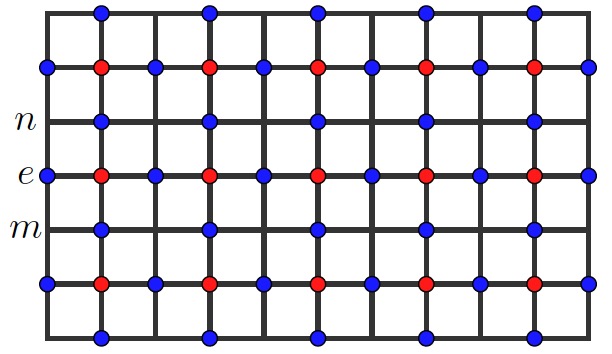} 
\captionof{figure}{ The body center square lattice where $\sigma$ spins colored in red live on the center  and $\tau$ spins colored in blue live on the link of each plaquette. }\label{(0,1)form}
\end{figure}

\begin{figure}[htpb] 
\centering 
\includegraphics[width=0.4\textwidth]{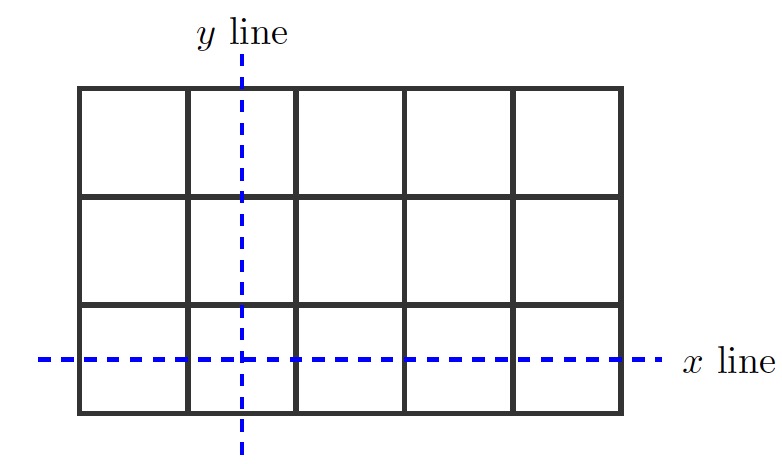} 
\captionof{figure}{Two generators of 1-form $\mathbb{Z}_2$ symmetry on torus. }\label{generators}
\end{figure}

For example, two generators of the cohomology group are spin flips on the link which crosses the $x$ line and the $y$ line shown in Fig.~\ref{generators}. Physically, they correspond to inserting a different $\pi$ flux through the two holes in the torus.

Moreover, the non-onsite duality operator is given by decorated domain wall construction \cite{yoshida2016topological}:
\begin{eqnarray}
U_{(0,1)}=&&\prod_{<a,b>}\exp\left[\frac{\pi i}{4}(1-\sigma^{z}_{a})(1-\tau^{z}_{b})\right].
\end{eqnarray}

To see the effect of this duality operator,  one can start from the trivial Hamiltonian	
\begin{eqnarray}\label{UDW}
H^{(0,1)}_{0}:=-\sum^L_{i,j=1}\left(\sigma^x_{i+\frac{1}{2},j+\frac{1}{2}}+\tau^x_{i+\frac{1}{2},j}+\tau^x_{i,j+\frac{1}{2}}\right)
\end{eqnarray}
and  the nontrivial SPT  Hamiltonian can be arrived at after conjugating this Hamiltonian by the duality operator:
\begin{eqnarray}
H^{(0,1)}_{1}=&& U_{(0,1)}H^{(0,1)}_{0}U_{(0,1)}^\dagger\nonumber\\ =&& -\sum^{L}_{i,j=1}\sigma^x_{i+\frac{1}{2},j+\frac{1}{2}}\tau^z_{i,j+\frac{1}{2}}\tau^z_{i+1,j+\frac{1}{2}}\tau^z_{i+\frac{1}{2},j}\tau^z_{i+\frac{1}{2},j+1}\nonumber\\&&-\sum^{L}_{i,j=1}(\tau^x_{i+\frac{1}{2},j}\sigma^z_{i+\frac{1}{2},j+\frac{1}{2}}\sigma^z_{i+\frac{1}{2},j-\frac{1}{2}}\nonumber\\&&+\tau^x_{i,j+\frac{1}{2}}\sigma^z_{i+\frac{1}{2},j+\frac{1}{2}}\sigma^z_{i-\frac{1}{2},j+\frac{1}{2}}).
\end{eqnarray}

Now let us start to discuss ingappabilities of self-dual systems. We can also consider the symmetry twisted boundary condition on the $x$ direction by 0-form $\mathbb{Z}_{2}$ symmetry:
\begin{eqnarray}\label{stbc1form}
\sigma_{i+L+\frac{1}{2},j+\frac{1}{2}}^{a}=U_G \sigma_{i+\frac{1}{2},j+\frac{1}{2}}^{a}U^{-1}_G, \text{with} \ i=1,\cdots,L,\nonumber\\~
\end{eqnarray}
where closed boundary is put between the vertical line $i=L$ and $i=1$. Then following the same calculation in section \ref{1dproof}, the twisted Hamiltonian is invariant under a modified duality symmetry 
\begin{eqnarray}
U^{(1)}_{(0,1)}= U_{(0,1)}\prod^{L}_{j=1}\tau^z_{1,j+\frac{1}{2}},
\end{eqnarray}
which anticommutes with 1-form $\mathbb{Z}_2$ symmetry generator on the $x$ direction. Thus all the eigenstates of the STBC
Hamiltonian are exactly doubly degenerate. Finally, we can arrive at the conclusion that the spectra of Hamiltonian under PBC must be either gapless or gapped with nontrivial ground state degeneracy, which finishes the proof of ingappabilities.

\section{Application and concrete examples} \label{sec 3}
In this section, we will introduce concrete examples besides the direct combination of different SPT Hamiltonians.  More precisely, we will discuss two examples in one dimension and one in two dimensions and show that their spectrum is either gapless or gapped with nontrivial ground state degeneracy.
\subsection{Self-dual model-1 in one dimension}
The first example is a Hamiltonian which combines the trivial and nontrivial 1d SPT Hamiltonian with a next-nearest neighbor (NNN) Ising interaction and it is also discussed in the reference \cite{Tantivasadakarn:2021wdv}:
\begin{eqnarray}\label{eq:NNN}
H=&&-\sum^L_{i=1}(\sigma^x_i+\tau^x_{i+\frac{1}{2}})-\sum_{i=1}^{L} (\sigma^z_{i} \tau_{i+\frac{1}{2}}^x \sigma^z_{i+1} + \tau_{i-\frac{1}{2}}^z\sigma^x_{i} \tau_{i+\frac{1}{2}}^z)\nonumber\\&&-J\sum^L_{i=1}(\sigma^z_i\sigma^z_{i+1}+\tau^z_{i+\frac{1}{2}}\tau^z_{i+\frac{3}{2}}).
\end{eqnarray}
When $J=0$, this Hamiltonian is described by a free boson CFT in the low energy \cite{suzuki1971relationship,keating2004random}:
\begin{eqnarray}
H=K\int (\partial_x\varphi)^2+\frac{1}{4K}\int (\partial_x\theta)^2
\end{eqnarray}
where $\varphi$ and $\theta$ are 2$\pi$ periodic field and satisfies commutation relations $[\partial_x \theta(x),\varphi(y)]=2\pi i\delta(x-y)$. And the Luttinger liquid parameter $K=\frac{1}{4}$.

The symmetries of our interest act as:
\begin{eqnarray}
&&U_A:\varphi\to -\varphi,~\theta\to -\theta,\qquad U_G:\varphi\to -\varphi,~\theta\to \pi-\theta,\nonumber\\
&&U_{(0,0)}: \varphi\to \varphi+\pi,~\theta\to \theta.
\end{eqnarray}
This CFT has a mixed anomaly with respect to these three symmetries \cite{metlitski2018intrinsic,yao2019anomaly}, which is consistent with the ingappability on the lattice.

When $J\ne0$, we can perform the Kramers-Wannier (KW) transformation\footnote{Here we treat $\tau$ and $\sigma$ as one kind of spin in Kramers-Wannier transformation.} to study the spectrum, which maps this Hamiltonian to an XXZ chain \cite{Tantivasadakarn:2021wdv}:
\begin{eqnarray}\label{eq:KW dual}
H=&&-\sum^L_{i=1}(\tau^z_{i-\frac{1}{2}}\sigma^z_i+\sigma^z_i\tau^z_{i+\frac{1}{2}})+\sum_{i=1}^{L} (\sigma^y_{i} \tau_{i+\frac{1}{2}}^y+ \tau_{i-\frac{1}{2}}^y\sigma^y_{i} )\nonumber\\&&-J\sum^L_{i=1}(\sigma^x_i\tau^x_{i+\frac{1}{2}}+\tau^x_{i+\frac{1}{2}}\sigma^x_{i+1}).
\end{eqnarray}
When $-1<J\leq 1$, this dual Hamiltonian is described by the free boson CFT in the low energy. When $J>1$ or $J<-1$,  the last nearest neighboured term dominates,  which induces the ferromagnetic/antiferromagnetic order of $\sigma^x$ and $\tau^x$ operators.  When $J=-1$, the gapped and gapless regimes are separated by a multicritical point with dynamical critical exponent $z=2$. 

Since the dynamical critical exponent and the center charge is invariant under the KW transformation, we can conclude that the Hamiltonian \eqref{eq:NNN} is described by a free boson CFT when $-1<J\leq 1$ while the multicritical point on $J=-1$  has dynamical critical exponent $z=2$. When $J>1$ or $J<-1$, since the last term in Eq.\eqref{eq:KW dual} is relevant, the NNN Ising interaction before KW transformation is also relevant, namely, it dominates the Hamiltonian \eqref{eq:NNN}. As a result, $\sigma^z$ and $\tau^z$ have nonzero expectation value and the $\mathbb{Z}^A_2\times\mathbb{Z}^G_2$ symmetry is 
spontaneously broken. 
\subsection{Self-dual model-2 in one dimension} 
The second example is a Hamiltonian which is a combination of another two cluster-like terms with a NNN Ising interaction \cite{Tantivasadakarn:2021wdv}:
\begin{eqnarray}\label{eq:NNN2}
H=&&-\sum^L_{i=1}(\tau^y_{i-\frac{1}{2}}\sigma^x_i \tau^y_{i+\frac{1}{2}}+\sigma^y_i\tau^x_{i+\frac{1}{2}}\sigma^y_{i+1})\nonumber\\&&-\sum_{i=1}^{L} (\sigma^z_{i-1} \tau_{i-\frac{1}{2}}^x\sigma^x_{i}\tau_{i+\frac{1}{2}}^x \sigma^z_{i+1} + \tau_{i-\frac{1}{2}}^z\sigma^x_{i} \tau_{i+\frac{1}{2}}^x \sigma^x_{i+1}\tau_{i+\frac{3}{2}}^z)\nonumber\\&&-J\sum^L_{i=1}(\sigma^z_i\sigma^z_{i+1}+\tau^z_{i+\frac{1}{2}}\tau^z_{i+\frac{3}{2}}).
\end{eqnarray}   
 The first and second terms also belong to the trivial and nontrivial SPT phases protected by $\mathbb{Z}^A_2 \times \mathbb{Z}^G_2$ symmetry\footnote{If we include the time-reversal symmetry $T=K$ which is complex conjugation, these two terms and the cluster chain \eqref{cluster} are distinct nontrivial SPT phases.}.   Indeed, these two terms are mapped to each other by the product of $\prod^L_{j=1}\sigma^z_j \tau^z_{j+\frac{1}{2}}$ and the original duality transformation $U_{(0,0)}$. Since  $\prod^L_{j=1}\sigma^z_j \tau^z_{j+\frac{1}{2}}$ is an onsite operator, it  does not affect the ingappability of the self-dual model above.

Now, we can perform the KW transformation to study the spectrum and this resulting Hamiltonian is:
\begin{eqnarray}\label{eq:KWdual2}
H=&&\sum^L_{i=1}(\tau^z_{i-\frac{1}{2}}\sigma^x_i \tau^x_{i+\frac{1}{2}}\sigma^z_{i+1}+\sigma^z_i\tau^x_{i+\frac{1}{2}}\sigma^x_{i+1}\tau^z_{i+\frac{3}{2}})\nonumber\\&&+\sum^L_{i=1}(\tau^y_{i-\frac{1}{2}}\sigma^x_i \tau^x_{i+\frac{1}{2}}\sigma^y_{i+1}+\sigma^y_i\tau^x_{i+\frac{1}{2}}\sigma^x_{i+1}\tau^y_{i+\frac{3}{2}})\nonumber\\&&-J\sum^L_{i=1}(\sigma^x_i\tau^x_{i+\frac{1}{2}}+\tau^x_{i+\frac{1}{2}}\sigma^x_{i+1}).
\end{eqnarray} 

To be more convenient, we label $\sigma$ and $\tau$ by one notation, namely, $\sigma^{x/z}_i=X_{2i}/Z_{2i}$ and $\tau^{x/z}_{i+\frac{1}{2}}=X_{2i+1}/Z_{2i+1}$. Then the Hamiltonian \eqref{eq:KWdual2} is rewritten as:
\begin{eqnarray}\label{eq:KWdual21}
H=&&\sum^{2L}_{i=1}(Z_{i-1}X_{i} X_{i+1}Z_{i+2}+Y_{i-1}X_{i} X_{i+1}Y_{i+2})\nonumber\\&&-J\sum^{2L}_{i=1}X_i X_{i+1}.
\end{eqnarray} 
The symmetry of our interest is the diagonal spin flip in the $x$ direction and three-site translation:
\begin{eqnarray}
&&R^{\pi}_{x}=\prod^{2L}_{j=1}\exp\left(\frac{\pi i}{2}(1-X_j)\right),\nonumber\\&& T_3 X_j T^{-1}_3=X_{j+3}, \qquad T_3 Z_j T^{-1}_3=Z_{j+3}.
\end{eqnarray}

We apply the Jordan-Wigner (JW) transformation which maps the spin operator to the fermion operator:
\begin{eqnarray}
X_{i}=(-1)^{n_i+1}=2f^{\dagger}_i f_i-1,\quad Z_i=\prod_{j=1}^{i-1}(-1)^{n_j}(f^{\dagger}_i+f_i) ,\nonumber\\~
\end{eqnarray}
 and the Hamiltonian \eqref{eq:KWdual21} is mapped to a fermion chain:
\begin{eqnarray}\label{eq:fermion}
H=&&2\sum^{2L}_{i=1}f^{\dagger}_{i}f_{i+3}+h.c-J\sum^{2L}_{i=1}(2n_i-1)(2 n_{i+1}-1).\nonumber\\~
\end{eqnarray} 
It is natural to consider the case $L\in 3\mathbb{Z}$. Then the first term is equivalent to three decoupled fermion chains with the nearest neighboured hopping term and the second term corresponds to a local interchain interaction \footnote{If the length $L$ is not a multiple of 3, this Hamiltonian is equivalent to the nearest neighboured hopping term with a long rang interaction when $J\ne 0$}.

Hence, when $J=0$, low energy theory is three decoupled Dirac fermions.
To discuss the spectra under interaction, we perform the standard Bosonization procedure and the low energy theory is mapped to three decoupled free boson CFTs:
\begin{eqnarray}
H_{LL}=\frac{1}{2\pi}\sum^3_{i=1}\int \left(K(\partial_x \varphi_i)^2+\frac{1}{4K}(\partial_x \theta_i)^2\right)dx
\end{eqnarray} 
where Luttinger liquid parameter $K=\frac{1}{4}$.

The dictionary relating the spin operators and effective low energy field operators is given by:
\begin{eqnarray}\label{eq:dictionary1}
&&X_{3k+j}\sim \frac{1}{2\pi}\partial_x \varphi_j-(-1)^{k}a_1 \cos\varphi_j, \nonumber\\
&&Z_{3k+j}+iY_{3k+j}\sim e^{i\theta_j}(b_0 (-1)^{k}+b_1\cos\varphi_j),
\end{eqnarray} 
where $a_1$, $b_0$ and $b_1$ are non-universal numbers. By this relation, we can obtain the action of symmetry of our interest in low energy:
\begin{eqnarray}
&&R^{\pi}_x: \theta_j \to \theta_j+\pi, \varphi_j\to \varphi_j, \nonumber\\&& T_3: \theta_j \to \theta_j+\pi, \varphi_j\to \varphi_j+\pi .
\end{eqnarray}

Moreover, the interchain interaction corresponds to two terms in the low energy:
\begin{eqnarray}
H^1_{int}=&&-\frac{J}{4\pi^2}\int (\partial_x\varphi_1 \partial_x\varphi_2+\partial_x\varphi_2 \partial_x\varphi_3+\partial_x\varphi_3 \partial_x\varphi_1)dx,\nonumber\\H^2_{int}=&&-Ja^2_1 \int (\cos\varphi_1\cos\varphi_2+\cos\varphi_2\cos\varphi_3\nonumber\\&&\qquad\quad-\cos\varphi_3\cos\varphi_1)dx\nonumber\\=&&-\frac{Ja^2_1}{2} \int [\cos(\varphi_1-\varphi_2)+\cos(\varphi_1+\varphi_2) \nonumber\\&&\qquad\quad+\cos(\varphi_2-\varphi_3)+\cos(\varphi_2+\varphi_3)\nonumber\\&&\qquad\quad-\cos(\varphi_3-\varphi_1)-\cos(\varphi_3+\varphi_1)]dx.
\end{eqnarray} 
We can diagonalize the first term  and three eigenvalues are $-1,-1,2$ times the 
coefficient $-\frac{J}{8\pi^2}$.  This term will modify Luttinger liquid parameter as follows:
\begin{eqnarray}
&&H_{LL}+H^1_{int}\nonumber\\=&&\frac{1}{2\pi}\sum^3_{i=1}\int \left(K^i_{eff}(\partial_x \varphi'_i)^2+\frac{1}{4K^i_{eff}}(\partial_x \theta'_i)^2\right)dx\nonumber\\~
\end{eqnarray} 
where $K^1_{eff}=K^2_{eff}=\frac{1}{4}(1+\frac{J}{\pi}), K^3_{eff}=\frac{1}{4}(1-\frac{2J}{\pi})$.
The relation between $\varphi$ and $\varphi'$ is $\varphi_i=\sum^3_{j=1}A_{ij} \varphi'_j$ where the matrix $A$ is:
\begin{equation}
\begin{aligned}
A=\left(\begin{array}{ccc}-0.7152 & 0.3938  & 0.5774 \\ 0.0166  & -0.8163 & 0.5774  \\ 0.6987 & 0.4225 & 0.5774 \end{array}\right) .
\end{aligned}
\end{equation}
We can calculate the scaling dimension of the term $\cos(\varphi_i\pm\varphi_{j})$ in the second low energy interaction:
\begin{eqnarray}
&&\frac{(A_{i1}\pm A_{j1})^2}{4K^1_{eff}}+\frac{(A_{i2}\pm A_{j2})^2}{4K^2_{eff}}+\frac{(A_{i3}\pm A_{j3})^2}{4K^3_{eff}}\nonumber\\=&&\frac{2}{4K^1_{eff}}+(A_{i3}\pm A_{j3})^2(\frac{1}{4K^3_{eff}}-\frac{1}{4K^1_{eff}}),
\end{eqnarray} 
where we use the fact that $A$ is an orthogonal matrix.

When $J$ is small, the scaling dimension can be simplified as
\begin{eqnarray}
 \cos(\varphi_i\pm\varphi_{j}):\quad  2\pm\frac{2J}{\pi}.
\end{eqnarray}
Therefore, the term $\cos(\varphi_1-\varphi_{2})+\cos(\varphi_2-\varphi_{3})-\cos(\varphi_3-\varphi_{1})$ is relevant when $J$ is positive. However,  $\varphi_1-\varphi_{2}$, $\varphi_2-\varphi_{3}$ and $\varphi_3-\varphi_{1}$  are linearly dependent\footnote{The configuration with lowest energy is $(\varphi_1-\varphi_{2},\varphi_2-\varphi_{3})=(\frac{\pi}{3},\frac{\pi}{3})/(-\frac{\pi}{3},-\frac{\pi}{3})$} and there is another independent degree of freedom $\varphi_1+\varphi_2+\varphi_3$ . Thus the low-energy theory after adding interaction is still gapless with center charge 1,  which implies that the Hamiltonian \eqref{eq:NNN2} before  KW transformation is also gapless with center charge 1.

On the other hand, while when $J$ is negative, $\cos(\varphi_1+\varphi_{2})+\cos(\varphi_2+\varphi_{3})-\cos(\varphi_3+\varphi_{1})$ is relevant. Then the ground states are gapped with configuration: $\varphi_1=\varphi_3=0, \varphi_2=\pi$ and $\varphi_1=\varphi_3=\pi, \varphi_2=0$ ,which break the translation symmetry but don't break the spin flip in the $x$ direction. Based on the properties of KW transformation, the Hamiltonian \eqref{eq:NNN2} is in the spontaneous
symmetry breaking (SSB) phase of diagonal spin flip symmetry $\prod^{L}_{i=1}\sigma^x_i\tau^x_{i+\frac{1}{2}}$.
\subsection{Self-dual model in two dimensions}
In two dimensions, we briefly introduce a spin model on the triangle lattice which has been studied in detail in \cite{PhysRevB.103.L140412,PhysRevB.103.144437,tantivasadakarn2021building}. The Hamiltonian is the combination of trivial and nontrivial SPT Hamiltonian protected by $(\mathbb{Z}_2)^3$ onsite symmetry with nearest-neighbor antiferromagnetic ($J>0$) Ising interactions within each of the three sublattices:

\begin{eqnarray}
H=&&H^{(0,0,0)}_0+H^{(0,0,0)}_1 +J \sum_{<v_1,v_2>}\sigma^z_{v_1}\sigma^z_{v_2}\nonumber\\&&+J \sum_{<v_1,v_2>}\tau^z_{v_1}\tau^z_{v_2}+J \sum_{<v_1,v_2>}\mu^z_{v_1}\mu^z_{v_2}.
\end{eqnarray}

The numerical calculation shows this model is in an FM phase breaking $(\mathbb{Z}_2)^3$ symmetry when $J$ is smaller than 0.42. On the other hand, this model is in a direct first-order transition between the SPT phases whose universality
is that of the SO(5) DQCP when $J$ is large. Moreover,  the  $\mathbb{Z}_2^{(0,0,0)}$ and ($\mathbb{Z}_2)^3  $ symmetry belong to a subgroup of SO(5) and there is a mixed 't Hooft anomaly between them in the low energy whose inflow action is a $\mathbb{Z}_2$ phase:
\begin{eqnarray}
i\pi A_1A_2A_3A_4.
\end{eqnarray}
Here $A_i$ is the background gauge field of $\mathbb{Z}_2^{(0,0,0)}$ duality and ($\mathbb{Z}_2)^3$ symmetry respective. This anomaly inflow action is consistent with the ingappability on the lattice.

\section{Conclusion and discussion}\label{sec 4}
 In this work, we focus on the quantum many-body systems which are self-dual under duality transformation connecting different SPTs.  Such self-duality often forces
the systems to be critical points separating duality-related phases. We prove
the ingappability of the bulk spectrum of these self-dual models based on the spectrum robustness argument on the STBC. As a direct result, these self-dual systems can be first order phase transition/SSB phase with nontrivial gapped ground state degeneracy or the continuous phase transition whose spectrum is gapless. This is equivalent to the statement that the
symmetry group at criticality has a mixed 't Hooft
anomaly.  We apply our method to several cases: 1. $d$ dimension self-dual systems with 0-form $(\mathbb{Z}_2)^{d+1}$ symmetry, 2. self-dual systems with $\mathbb{Z}_2 \times \mathbb{Z}_2$ line symmetry in arbitrary spatial dimensions, 3. two dimensional self-dual systems with $\mathbb{Z}_2$ 0-form and  $\mathbb{Z}_2$ 1-form symmetry. Moreover, we illustrate this result with several examples in one and two dimensions, whose spectrum can be obtained by analytical or numerical calculations. 

 For future studies, one important question is how to generalize our method to the critical points between fermionic SPT phases. An interesting example of such a fermionic duality transformation is the Majorana translation operator, which connects the trivial and nontrivial Majorana chains. It has been shown that there is an ingappability of the self-dual model under such duality transformation \cite{PhysRevLett.117.166802}. Moreover, it is also quite interesting to study the geometric description of ingappabilities of self-dual systems separating SPTs protected by anti-unitary (time-reversal) symmetries or crystal symmetries.

\section*{Acknowledgements}

L. L. is supported by Global Science Graduate Course
(GSGC) program at the University of Tokyo.

\bibliography{bib}

\begin{thebibliography}{79}%
\makeatletter
\providecommand \@ifxundefined [1]{%
 \@ifx{#1\undefined}
}%
\providecommand \@ifnum [1]{%
 \ifnum #1\expandafter \@firstoftwo
 \else \expandafter \@secondoftwo
 \fi
}%
\providecommand \@ifx [1]{%
 \ifx #1\expandafter \@firstoftwo
 \else \expandafter \@secondoftwo
 \fi
}%
\providecommand \natexlab [1]{#1}%
\providecommand \enquote  [1]{``#1''}%
\providecommand \bibnamefont  [1]{#1}%
\providecommand \bibfnamefont [1]{#1}%
\providecommand \citenamefont [1]{#1}%
\providecommand \href@noop [0]{\@secondoftwo}%
\providecommand \href [0]{\begingroup \@sanitize@url \@href}%
\providecommand \@href[1]{\@@startlink{#1}\@@href}%
\providecommand \@@href[1]{\endgroup#1\@@endlink}%
\providecommand \@sanitize@url [0]{\catcode `\\12\catcode `\$12\catcode
  `\&12\catcode `\#12\catcode `\^12\catcode `\_12\catcode `\%12\relax}%
\providecommand \@@startlink[1]{}%
\providecommand \@@endlink[0]{}%
\providecommand \url  [0]{\begingroup\@sanitize@url \@url }%
\providecommand \@url [1]{\endgroup\@href {#1}{\urlprefix }}%
\providecommand \urlprefix  [0]{URL }%
\providecommand \Eprint [0]{\href }%
\providecommand \doibase [0]{http://dx.doi.org/}%
\providecommand \selectlanguage [0]{\@gobble}%
\providecommand \bibinfo  [0]{\@secondoftwo}%
\providecommand \bibfield  [0]{\@secondoftwo}%
\providecommand \translation [1]{[#1]}%
\providecommand \BibitemOpen [0]{}%
\providecommand \bibitemStop [0]{}%
\providecommand \bibitemNoStop [0]{.\EOS\space}%
\providecommand \EOS [0]{\spacefactor3000\relax}%
\providecommand \BibitemShut  [1]{\csname bibitem#1\endcsname}%
\let\auto@bib@innerbib\@empty
\bibitem [{\citenamefont {Haldane}(1983)}]{HALDANE1983464}%
  \BibitemOpen
  \bibfield  {author} {\bibinfo {author} {\bibfnamefont {F.}~\bibnamefont
  {Haldane}},\ }\href {\doibase https://doi.org/10.1016/0375-9601(83)90631-X}
  {\bibfield  {journal} {\bibinfo  {journal} {Physics Letters A}\ }\textbf
  {\bibinfo {volume} {93}},\ \bibinfo {pages} {464} (\bibinfo {year}
  {1983})}\BibitemShut {NoStop}%
\bibitem [{\citenamefont {Affleck}\ \emph {et~al.}(1987)\citenamefont
  {Affleck}, \citenamefont {Kennedy}, \citenamefont {Lieb},\ and\ \citenamefont
  {Tasaki}}]{PhysRevLett.59.799}%
  \BibitemOpen
  \bibfield  {author} {\bibinfo {author} {\bibfnamefont {I.}~\bibnamefont
  {Affleck}}, \bibinfo {author} {\bibfnamefont {T.}~\bibnamefont {Kennedy}},
  \bibinfo {author} {\bibfnamefont {E.~H.}\ \bibnamefont {Lieb}}, \ and\
  \bibinfo {author} {\bibfnamefont {H.}~\bibnamefont {Tasaki}},\ }\href
  {\doibase 10.1103/PhysRevLett.59.799} {\bibfield  {journal} {\bibinfo
  {journal} {Phys. Rev. Lett.}\ }\textbf {\bibinfo {volume} {59}},\ \bibinfo
  {pages} {799} (\bibinfo {year} {1987})}\BibitemShut {NoStop}%
\bibitem [{\citenamefont {Fidkowski}\ and\ \citenamefont
  {Kitaev}(2011)}]{PhysRevB.83.075103}%
  \BibitemOpen
  \bibfield  {author} {\bibinfo {author} {\bibfnamefont {L.}~\bibnamefont
  {Fidkowski}}\ and\ \bibinfo {author} {\bibfnamefont {A.}~\bibnamefont
  {Kitaev}},\ }\href {\doibase 10.1103/PhysRevB.83.075103} {\bibfield
  {journal} {\bibinfo  {journal} {Phys. Rev. B}\ }\textbf {\bibinfo {volume}
  {83}},\ \bibinfo {pages} {075103} (\bibinfo {year} {2011})}\BibitemShut
  {NoStop}%
\bibitem [{\citenamefont {Schuch}\ \emph {et~al.}(2011)\citenamefont {Schuch},
  \citenamefont {P\'erez-Garc\'{\i}a},\ and\ \citenamefont
  {Cirac}}]{PhysRevB.84.165139}%
  \BibitemOpen
  \bibfield  {author} {\bibinfo {author} {\bibfnamefont {N.}~\bibnamefont
  {Schuch}}, \bibinfo {author} {\bibfnamefont {D.}~\bibnamefont
  {P\'erez-Garc\'{\i}a}}, \ and\ \bibinfo {author} {\bibfnamefont
  {I.}~\bibnamefont {Cirac}},\ }\href {\doibase 10.1103/PhysRevB.84.165139}
  {\bibfield  {journal} {\bibinfo  {journal} {Phys. Rev. B}\ }\textbf {\bibinfo
  {volume} {84}},\ \bibinfo {pages} {165139} (\bibinfo {year}
  {2011})}\BibitemShut {NoStop}%
\bibitem [{\citenamefont {Pollmann}\ \emph {et~al.}(2010)\citenamefont
  {Pollmann}, \citenamefont {Turner}, \citenamefont {Berg},\ and\ \citenamefont
  {Oshikawa}}]{PhysRevB.81.064439}%
  \BibitemOpen
  \bibfield  {author} {\bibinfo {author} {\bibfnamefont {F.}~\bibnamefont
  {Pollmann}}, \bibinfo {author} {\bibfnamefont {A.~M.}\ \bibnamefont
  {Turner}}, \bibinfo {author} {\bibfnamefont {E.}~\bibnamefont {Berg}}, \ and\
  \bibinfo {author} {\bibfnamefont {M.}~\bibnamefont {Oshikawa}},\ }\href
  {\doibase 10.1103/PhysRevB.81.064439} {\bibfield  {journal} {\bibinfo
  {journal} {Phys. Rev. B}\ }\textbf {\bibinfo {volume} {81}},\ \bibinfo
  {pages} {064439} (\bibinfo {year} {2010})}\BibitemShut {NoStop}%
\bibitem [{\citenamefont {Chen}\ \emph {et~al.}(2011)\citenamefont {Chen},
  \citenamefont {Gu},\ and\ \citenamefont {Wen}}]{PhysRevB.83.035107}%
  \BibitemOpen
  \bibfield  {author} {\bibinfo {author} {\bibfnamefont {X.}~\bibnamefont
  {Chen}}, \bibinfo {author} {\bibfnamefont {Z.-C.}\ \bibnamefont {Gu}}, \ and\
  \bibinfo {author} {\bibfnamefont {X.-G.}\ \bibnamefont {Wen}},\ }\href
  {\doibase 10.1103/PhysRevB.83.035107} {\bibfield  {journal} {\bibinfo
  {journal} {Phys. Rev. B}\ }\textbf {\bibinfo {volume} {83}},\ \bibinfo
  {pages} {035107} (\bibinfo {year} {2011})}\BibitemShut {NoStop}%
\bibitem [{\citenamefont {Chen}\ \emph {et~al.}(2012)\citenamefont {Chen},
  \citenamefont {Gu}, \citenamefont {Liu},\ and\ \citenamefont
  {Wen}}]{chen2012symmetry}%
  \BibitemOpen
  \bibfield  {author} {\bibinfo {author} {\bibfnamefont {X.}~\bibnamefont
  {Chen}}, \bibinfo {author} {\bibfnamefont {Z.-C.}\ \bibnamefont {Gu}},
  \bibinfo {author} {\bibfnamefont {Z.-X.}\ \bibnamefont {Liu}}, \ and\
  \bibinfo {author} {\bibfnamefont {X.-G.}\ \bibnamefont {Wen}},\ }\href@noop
  {} {\bibfield  {journal} {\bibinfo  {journal} {Science}\ }\textbf {\bibinfo
  {volume} {338}},\ \bibinfo {pages} {1604} (\bibinfo {year}
  {2012})}\BibitemShut {NoStop}%
\bibitem [{\citenamefont {Vishwanath}\ and\ \citenamefont
  {Senthil}(2013)}]{PhysRevX.3.011016}%
  \BibitemOpen
  \bibfield  {author} {\bibinfo {author} {\bibfnamefont {A.}~\bibnamefont
  {Vishwanath}}\ and\ \bibinfo {author} {\bibfnamefont {T.}~\bibnamefont
  {Senthil}},\ }\href {\doibase 10.1103/PhysRevX.3.011016} {\bibfield
  {journal} {\bibinfo  {journal} {Phys. Rev. X}\ }\textbf {\bibinfo {volume}
  {3}},\ \bibinfo {pages} {011016} (\bibinfo {year} {2013})}\BibitemShut
  {NoStop}%
\bibitem [{\citenamefont {Chen}\ \emph {et~al.}(2010)\citenamefont {Chen},
  \citenamefont {Gu},\ and\ \citenamefont {Wen}}]{PhysRevB.82.155138}%
  \BibitemOpen
  \bibfield  {author} {\bibinfo {author} {\bibfnamefont {X.}~\bibnamefont
  {Chen}}, \bibinfo {author} {\bibfnamefont {Z.-C.}\ \bibnamefont {Gu}}, \ and\
  \bibinfo {author} {\bibfnamefont {X.-G.}\ \bibnamefont {Wen}},\ }\href
  {\doibase 10.1103/PhysRevB.82.155138} {\bibfield  {journal} {\bibinfo
  {journal} {Phys. Rev. B}\ }\textbf {\bibinfo {volume} {82}},\ \bibinfo
  {pages} {155138} (\bibinfo {year} {2010})}\BibitemShut {NoStop}%
\bibitem [{\citenamefont {Kennedy}\ and\ \citenamefont
  {Tasaki}(1992{\natexlab{a}})}]{PhysRevB.45.304}%
  \BibitemOpen
  \bibfield  {author} {\bibinfo {author} {\bibfnamefont {T.}~\bibnamefont
  {Kennedy}}\ and\ \bibinfo {author} {\bibfnamefont {H.}~\bibnamefont
  {Tasaki}},\ }\href {\doibase 10.1103/PhysRevB.45.304} {\bibfield  {journal}
  {\bibinfo  {journal} {Phys. Rev. B}\ }\textbf {\bibinfo {volume} {45}},\
  \bibinfo {pages} {304} (\bibinfo {year} {1992}{\natexlab{a}})}\BibitemShut
  {NoStop}%
\bibitem [{\citenamefont {Kennedy}\ and\ \citenamefont
  {Tasaki}(1992{\natexlab{b}})}]{kennedy1992hidden}%
  \BibitemOpen
  \bibfield  {author} {\bibinfo {author} {\bibfnamefont {T.}~\bibnamefont
  {Kennedy}}\ and\ \bibinfo {author} {\bibfnamefont {H.}~\bibnamefont
  {Tasaki}},\ }\href@noop {} {\bibfield  {journal} {\bibinfo  {journal}
  {Communications in mathematical physics}\ }\textbf {\bibinfo {volume}
  {147}},\ \bibinfo {pages} {431} (\bibinfo {year}
  {1992}{\natexlab{b}})}\BibitemShut {NoStop}%
\bibitem [{\citenamefont {Oshikawa}(1992)}]{oshikawa1992hidden}%
  \BibitemOpen
  \bibfield  {author} {\bibinfo {author} {\bibfnamefont {M.}~\bibnamefont
  {Oshikawa}},\ }\href@noop {} {\bibfield  {journal} {\bibinfo  {journal}
  {Journal of Physics: Condensed Matter}\ }\textbf {\bibinfo {volume} {4}},\
  \bibinfo {pages} {7469} (\bibinfo {year} {1992})}\BibitemShut {NoStop}%
\bibitem [{\citenamefont {Pollmann}\ \emph {et~al.}(2012)\citenamefont
  {Pollmann}, \citenamefont {Berg}, \citenamefont {Turner},\ and\ \citenamefont
  {Oshikawa}}]{PhysRevB.85.075125}%
  \BibitemOpen
  \bibfield  {author} {\bibinfo {author} {\bibfnamefont {F.}~\bibnamefont
  {Pollmann}}, \bibinfo {author} {\bibfnamefont {E.}~\bibnamefont {Berg}},
  \bibinfo {author} {\bibfnamefont {A.~M.}\ \bibnamefont {Turner}}, \ and\
  \bibinfo {author} {\bibfnamefont {M.}~\bibnamefont {Oshikawa}},\ }\href
  {\doibase 10.1103/PhysRevB.85.075125} {\bibfield  {journal} {\bibinfo
  {journal} {Phys. Rev. B}\ }\textbf {\bibinfo {volume} {85}},\ \bibinfo
  {pages} {075125} (\bibinfo {year} {2012})}\BibitemShut {NoStop}%
\bibitem [{\citenamefont {Else}\ and\ \citenamefont
  {Nayak}(2014)}]{PhysRevB.90.235137}%
  \BibitemOpen
  \bibfield  {author} {\bibinfo {author} {\bibfnamefont {D.~V.}\ \bibnamefont
  {Else}}\ and\ \bibinfo {author} {\bibfnamefont {C.}~\bibnamefont {Nayak}},\
  }\href {\doibase 10.1103/PhysRevB.90.235137} {\bibfield  {journal} {\bibinfo
  {journal} {Phys. Rev. B}\ }\textbf {\bibinfo {volume} {90}},\ \bibinfo
  {pages} {235137} (\bibinfo {year} {2014})}\BibitemShut {NoStop}%
\bibitem [{\citenamefont {Chen}\ \emph {et~al.}(2014)\citenamefont {Chen},
  \citenamefont {Lu},\ and\ \citenamefont
  {Vishwanath}}]{Chen2013SymmetryprotectedTP}%
  \BibitemOpen
  \bibfield  {author} {\bibinfo {author} {\bibfnamefont {X.}~\bibnamefont
  {Chen}}, \bibinfo {author} {\bibfnamefont {Y.-M.}\ \bibnamefont {Lu}}, \ and\
  \bibinfo {author} {\bibfnamefont {A.}~\bibnamefont {Vishwanath}},\
  }\href@noop {} {\bibfield  {journal} {\bibinfo  {journal} {Nature
  communications}\ }\textbf {\bibinfo {volume} {5}},\ \bibinfo {pages} {3507}
  (\bibinfo {year} {2014})}\BibitemShut {NoStop}%
\bibitem [{\citenamefont {Wang}\ \emph
  {et~al.}(2021{\natexlab{a}})\citenamefont {Wang}, \citenamefont {Ning},\ and\
  \citenamefont {Cheng}}]{wang2021domain}%
  \BibitemOpen
  \bibfield  {author} {\bibinfo {author} {\bibfnamefont {Q.-R.}\ \bibnamefont
  {Wang}}, \bibinfo {author} {\bibfnamefont {S.-Q.}\ \bibnamefont {Ning}}, \
  and\ \bibinfo {author} {\bibfnamefont {M.}~\bibnamefont {Cheng}},\
  }\href@noop {} {\bibfield  {journal} {\bibinfo  {journal} {arXiv preprint
  arXiv:2104.13233}\ } (\bibinfo {year} {2021}{\natexlab{a}})}\BibitemShut
  {NoStop}%
\bibitem [{\citenamefont {Yoshida}(2016)}]{yoshida2016topological}%
  \BibitemOpen
  \bibfield  {author} {\bibinfo {author} {\bibfnamefont {B.}~\bibnamefont
  {Yoshida}},\ }\href@noop {} {\bibfield  {journal} {\bibinfo  {journal}
  {Physical Review B}\ }\textbf {\bibinfo {volume} {93}},\ \bibinfo {pages}
  {155131} (\bibinfo {year} {2016})}\BibitemShut {NoStop}%
\bibitem [{\citenamefont {Motrunich}\ and\ \citenamefont
  {Vishwanath}(2004)}]{PhysRevB.70.075104}%
  \BibitemOpen
  \bibfield  {author} {\bibinfo {author} {\bibfnamefont {O.~I.}\ \bibnamefont
  {Motrunich}}\ and\ \bibinfo {author} {\bibfnamefont {A.}~\bibnamefont
  {Vishwanath}},\ }\href {\doibase 10.1103/PhysRevB.70.075104} {\bibfield
  {journal} {\bibinfo  {journal} {Phys. Rev. B}\ }\textbf {\bibinfo {volume}
  {70}},\ \bibinfo {pages} {075104} (\bibinfo {year} {2004})}\BibitemShut
  {NoStop}%
\bibitem [{\citenamefont {You}\ \emph {et~al.}(2016)\citenamefont {You},
  \citenamefont {Bi}, \citenamefont {Mao},\ and\ \citenamefont
  {Xu}}]{PhysRevB.93.125101}%
  \BibitemOpen
  \bibfield  {author} {\bibinfo {author} {\bibfnamefont {Y.-Z.}\ \bibnamefont
  {You}}, \bibinfo {author} {\bibfnamefont {Z.}~\bibnamefont {Bi}}, \bibinfo
  {author} {\bibfnamefont {D.}~\bibnamefont {Mao}}, \ and\ \bibinfo {author}
  {\bibfnamefont {C.}~\bibnamefont {Xu}},\ }\href {\doibase
  10.1103/PhysRevB.93.125101} {\bibfield  {journal} {\bibinfo  {journal} {Phys.
  Rev. B}\ }\textbf {\bibinfo {volume} {93}},\ \bibinfo {pages} {125101}
  (\bibinfo {year} {2016})}\BibitemShut {NoStop}%
\bibitem [{\citenamefont {You}\ and\ \citenamefont
  {You}(2016)}]{PhysRevB.93.195141}%
  \BibitemOpen
  \bibfield  {author} {\bibinfo {author} {\bibfnamefont {Y.}~\bibnamefont
  {You}}\ and\ \bibinfo {author} {\bibfnamefont {Y.-Z.}\ \bibnamefont {You}},\
  }\href {\doibase 10.1103/PhysRevB.93.195141} {\bibfield  {journal} {\bibinfo
  {journal} {Phys. Rev. B}\ }\textbf {\bibinfo {volume} {93}},\ \bibinfo
  {pages} {195141} (\bibinfo {year} {2016})}\BibitemShut {NoStop}%
\bibitem [{\citenamefont {Jian}\ \emph {et~al.}(2018)\citenamefont {Jian},
  \citenamefont {Thomson}, \citenamefont {Rasmussen}, \citenamefont {Bi},\ and\
  \citenamefont {Xu}}]{PhysRevB.97.195115}%
  \BibitemOpen
  \bibfield  {author} {\bibinfo {author} {\bibfnamefont {C.-M.}\ \bibnamefont
  {Jian}}, \bibinfo {author} {\bibfnamefont {A.}~\bibnamefont {Thomson}},
  \bibinfo {author} {\bibfnamefont {A.}~\bibnamefont {Rasmussen}}, \bibinfo
  {author} {\bibfnamefont {Z.}~\bibnamefont {Bi}}, \ and\ \bibinfo {author}
  {\bibfnamefont {C.}~\bibnamefont {Xu}},\ }\href {\doibase
  10.1103/PhysRevB.97.195115} {\bibfield  {journal} {\bibinfo  {journal} {Phys.
  Rev. B}\ }\textbf {\bibinfo {volume} {97}},\ \bibinfo {pages} {195115}
  (\bibinfo {year} {2018})}\BibitemShut {NoStop}%
\bibitem [{\citenamefont {Senthil}\ \emph {et~al.}(2019)\citenamefont
  {Senthil}, \citenamefont {Son}, \citenamefont {Wang},\ and\ \citenamefont
  {Xu}}]{senthil2019duality}%
  \BibitemOpen
  \bibfield  {author} {\bibinfo {author} {\bibfnamefont {T.}~\bibnamefont
  {Senthil}}, \bibinfo {author} {\bibfnamefont {D.~T.}\ \bibnamefont {Son}},
  \bibinfo {author} {\bibfnamefont {C.}~\bibnamefont {Wang}}, \ and\ \bibinfo
  {author} {\bibfnamefont {C.}~\bibnamefont {Xu}},\ }\href@noop {} {\bibfield
  {journal} {\bibinfo  {journal} {Physics Reports}\ }\textbf {\bibinfo {volume}
  {827}},\ \bibinfo {pages} {1} (\bibinfo {year} {2019})}\BibitemShut {NoStop}%
\bibitem [{\citenamefont {Bi}\ and\ \citenamefont
  {Senthil}(2019)}]{PhysRevX.9.021034}%
  \BibitemOpen
  \bibfield  {author} {\bibinfo {author} {\bibfnamefont {Z.}~\bibnamefont
  {Bi}}\ and\ \bibinfo {author} {\bibfnamefont {T.}~\bibnamefont {Senthil}},\
  }\href {\doibase 10.1103/PhysRevX.9.021034} {\bibfield  {journal} {\bibinfo
  {journal} {Phys. Rev. X}\ }\textbf {\bibinfo {volume} {9}},\ \bibinfo {pages}
  {021034} (\bibinfo {year} {2019})}\BibitemShut {NoStop}%
\bibitem [{\citenamefont {Grover}\ and\ \citenamefont
  {Vishwanath}(2013)}]{PhysRevB.87.045129}%
  \BibitemOpen
  \bibfield  {author} {\bibinfo {author} {\bibfnamefont {T.}~\bibnamefont
  {Grover}}\ and\ \bibinfo {author} {\bibfnamefont {A.}~\bibnamefont
  {Vishwanath}},\ }\href {\doibase 10.1103/PhysRevB.87.045129} {\bibfield
  {journal} {\bibinfo  {journal} {Phys. Rev. B}\ }\textbf {\bibinfo {volume}
  {87}},\ \bibinfo {pages} {045129} (\bibinfo {year} {2013})}\BibitemShut
  {NoStop}%
\bibitem [{\citenamefont {Lu}\ and\ \citenamefont
  {Lee}(2014)}]{PhysRevB.89.195143}%
  \BibitemOpen
  \bibfield  {author} {\bibinfo {author} {\bibfnamefont {Y.-M.}\ \bibnamefont
  {Lu}}\ and\ \bibinfo {author} {\bibfnamefont {D.-H.}\ \bibnamefont {Lee}},\
  }\href {\doibase 10.1103/PhysRevB.89.195143} {\bibfield  {journal} {\bibinfo
  {journal} {Phys. Rev. B}\ }\textbf {\bibinfo {volume} {89}},\ \bibinfo
  {pages} {195143} (\bibinfo {year} {2014})}\BibitemShut {NoStop}%
\bibitem [{\citenamefont {Pixley}\ \emph {et~al.}(2014)\citenamefont {Pixley},
  \citenamefont {Shashi},\ and\ \citenamefont
  {Nevidomskyy}}]{PhysRevB.90.214426}%
  \BibitemOpen
  \bibfield  {author} {\bibinfo {author} {\bibfnamefont {J.~H.}\ \bibnamefont
  {Pixley}}, \bibinfo {author} {\bibfnamefont {A.}~\bibnamefont {Shashi}}, \
  and\ \bibinfo {author} {\bibfnamefont {A.~H.}\ \bibnamefont {Nevidomskyy}},\
  }\href {\doibase 10.1103/PhysRevB.90.214426} {\bibfield  {journal} {\bibinfo
  {journal} {Phys. Rev. B}\ }\textbf {\bibinfo {volume} {90}},\ \bibinfo
  {pages} {214426} (\bibinfo {year} {2014})}\BibitemShut {NoStop}%
\bibitem [{\citenamefont {Lahtinen}\ and\ \citenamefont
  {Ardonne}(2015)}]{PhysRevLett.115.237203}%
  \BibitemOpen
  \bibfield  {author} {\bibinfo {author} {\bibfnamefont {V.}~\bibnamefont
  {Lahtinen}}\ and\ \bibinfo {author} {\bibfnamefont {E.}~\bibnamefont
  {Ardonne}},\ }\href {\doibase 10.1103/PhysRevLett.115.237203} {\bibfield
  {journal} {\bibinfo  {journal} {Phys. Rev. Lett.}\ }\textbf {\bibinfo
  {volume} {115}},\ \bibinfo {pages} {237203} (\bibinfo {year}
  {2015})}\BibitemShut {NoStop}%
\bibitem [{\citenamefont {You}\ \emph {et~al.}(2018{\natexlab{a}})\citenamefont
  {You}, \citenamefont {He}, \citenamefont {Vishwanath},\ and\ \citenamefont
  {Xu}}]{PhysRevB.97.125112}%
  \BibitemOpen
  \bibfield  {author} {\bibinfo {author} {\bibfnamefont {Y.-Z.}\ \bibnamefont
  {You}}, \bibinfo {author} {\bibfnamefont {Y.-C.}\ \bibnamefont {He}},
  \bibinfo {author} {\bibfnamefont {A.}~\bibnamefont {Vishwanath}}, \ and\
  \bibinfo {author} {\bibfnamefont {C.}~\bibnamefont {Xu}},\ }\href {\doibase
  10.1103/PhysRevB.97.125112} {\bibfield  {journal} {\bibinfo  {journal} {Phys.
  Rev. B}\ }\textbf {\bibinfo {volume} {97}},\ \bibinfo {pages} {125112}
  (\bibinfo {year} {2018}{\natexlab{a}})}\BibitemShut {NoStop}%
\bibitem [{\citenamefont {Verresen}\ \emph {et~al.}(2017)\citenamefont
  {Verresen}, \citenamefont {Moessner},\ and\ \citenamefont
  {Pollmann}}]{PhysRevB.96.165124}%
  \BibitemOpen
  \bibfield  {author} {\bibinfo {author} {\bibfnamefont {R.}~\bibnamefont
  {Verresen}}, \bibinfo {author} {\bibfnamefont {R.}~\bibnamefont {Moessner}},
  \ and\ \bibinfo {author} {\bibfnamefont {F.}~\bibnamefont {Pollmann}},\
  }\href {\doibase 10.1103/PhysRevB.96.165124} {\bibfield  {journal} {\bibinfo
  {journal} {Phys. Rev. B}\ }\textbf {\bibinfo {volume} {96}},\ \bibinfo
  {pages} {165124} (\bibinfo {year} {2017})}\BibitemShut {NoStop}%
\bibitem [{\citenamefont {Tantivasadakarn}\ \emph
  {et~al.}(2021{\natexlab{a}})\citenamefont {Tantivasadakarn}, \citenamefont
  {Thorngren}, \citenamefont {Vishwanath},\ and\ \citenamefont
  {Verresen}}]{tantivasadakarn2021building}%
  \BibitemOpen
  \bibfield  {author} {\bibinfo {author} {\bibfnamefont {N.}~\bibnamefont
  {Tantivasadakarn}}, \bibinfo {author} {\bibfnamefont {R.}~\bibnamefont
  {Thorngren}}, \bibinfo {author} {\bibfnamefont {A.}~\bibnamefont
  {Vishwanath}}, \ and\ \bibinfo {author} {\bibfnamefont {R.}~\bibnamefont
  {Verresen}},\ }\href@noop {} {\bibfield  {journal} {\bibinfo  {journal}
  {arXiv preprint arXiv:2110.09512}\ } (\bibinfo {year}
  {2021}{\natexlab{a}})}\BibitemShut {NoStop}%
\bibitem [{\citenamefont {Mudry}\ \emph {et~al.}(2019)\citenamefont {Mudry},
  \citenamefont {Furusaki}, \citenamefont {Morimoto},\ and\ \citenamefont
  {Hikihara}}]{mudry2019quantum}%
  \BibitemOpen
  \bibfield  {author} {\bibinfo {author} {\bibfnamefont {C.}~\bibnamefont
  {Mudry}}, \bibinfo {author} {\bibfnamefont {A.}~\bibnamefont {Furusaki}},
  \bibinfo {author} {\bibfnamefont {T.}~\bibnamefont {Morimoto}}, \ and\
  \bibinfo {author} {\bibfnamefont {T.}~\bibnamefont {Hikihara}},\ }\href@noop
  {} {\bibfield  {journal} {\bibinfo  {journal} {Physical Review B}\ }\textbf
  {\bibinfo {volume} {99}},\ \bibinfo {pages} {205153} (\bibinfo {year}
  {2019})}\BibitemShut {NoStop}%
\bibitem [{\citenamefont {Aksoy}\ \emph
  {et~al.}(2021{\natexlab{a}})\citenamefont {Aksoy}, \citenamefont {Chen},
  \citenamefont {Ryu}, \citenamefont {Furusaki},\ and\ \citenamefont
  {Mudry}}]{aksoy2021stability}%
  \BibitemOpen
  \bibfield  {author} {\bibinfo {author} {\bibfnamefont {{\"O}.~M.}\
  \bibnamefont {Aksoy}}, \bibinfo {author} {\bibfnamefont {J.-H.}\ \bibnamefont
  {Chen}}, \bibinfo {author} {\bibfnamefont {S.}~\bibnamefont {Ryu}}, \bibinfo
  {author} {\bibfnamefont {A.}~\bibnamefont {Furusaki}}, \ and\ \bibinfo
  {author} {\bibfnamefont {C.}~\bibnamefont {Mudry}},\ }\href@noop {}
  {\bibfield  {journal} {\bibinfo  {journal} {Physical Review B}\ }\textbf
  {\bibinfo {volume} {103}},\ \bibinfo {pages} {205121} (\bibinfo {year}
  {2021}{\natexlab{a}})}\BibitemShut {NoStop}%
\bibitem [{\citenamefont {Kapustin}\ and\ \citenamefont
  {Thorngren}(2017)}]{Kapustin2017}%
  \BibitemOpen
  \bibfield  {author} {\bibinfo {author} {\bibfnamefont {A.}~\bibnamefont
  {Kapustin}}\ and\ \bibinfo {author} {\bibfnamefont {R.}~\bibnamefont
  {Thorngren}},\ }\enquote {\bibinfo {title} {Higher symmetry and gapped phases
  of gauge theories},}\ in\ \href {\doibase 10.1007/978-3-319-59939-7_5} {\emph
  {\bibinfo {booktitle} {Algebra, Geometry, and Physics in the 21st Century:
  Kontsevich Festschrift}}},\ \bibinfo {editor} {edited by\ \bibinfo {editor}
  {\bibfnamefont {D.}~\bibnamefont {Auroux}}, \bibinfo {editor} {\bibfnamefont
  {L.}~\bibnamefont {Katzarkov}}, \bibinfo {editor} {\bibfnamefont
  {T.}~\bibnamefont {Pantev}}, \bibinfo {editor} {\bibfnamefont
  {Y.}~\bibnamefont {Soibelman}}, \ and\ \bibinfo {editor} {\bibfnamefont
  {Y.}~\bibnamefont {Tschinkel}}}\ (\bibinfo  {publisher} {Springer
  International Publishing},\ \bibinfo {address} {Cham},\ \bibinfo {year}
  {2017})\ pp.\ \bibinfo {pages} {177--202}\BibitemShut {NoStop}%
\bibitem [{\citenamefont {Kapustin}\ and\ \citenamefont
  {Seiberg}(2014)}]{kapustin2014coupling}%
  \BibitemOpen
  \bibfield  {author} {\bibinfo {author} {\bibfnamefont {A.}~\bibnamefont
  {Kapustin}}\ and\ \bibinfo {author} {\bibfnamefont {N.}~\bibnamefont
  {Seiberg}},\ }\href@noop {} {\bibfield  {journal} {\bibinfo  {journal}
  {Journal of High Energy Physics}\ }\textbf {\bibinfo {volume} {2014}},\
  \bibinfo {pages} {1} (\bibinfo {year} {2014})}\BibitemShut {NoStop}%
\bibitem [{\citenamefont {Gaiotto}\ \emph {et~al.}(2015)\citenamefont
  {Gaiotto}, \citenamefont {Kapustin}, \citenamefont {Seiberg},\ and\
  \citenamefont {Willett}}]{Gaiotto:2014kfa}%
  \BibitemOpen
  \bibfield  {author} {\bibinfo {author} {\bibfnamefont {D.}~\bibnamefont
  {Gaiotto}}, \bibinfo {author} {\bibfnamefont {A.}~\bibnamefont {Kapustin}},
  \bibinfo {author} {\bibfnamefont {N.}~\bibnamefont {Seiberg}}, \ and\
  \bibinfo {author} {\bibfnamefont {B.}~\bibnamefont {Willett}},\ }\href
  {\doibase 10.1007/JHEP02(2015)172} {\bibfield  {journal} {\bibinfo  {journal}
  {JHEP}\ }\textbf {\bibinfo {volume} {02}},\ \bibinfo {pages} {172} (\bibinfo
  {year} {2015})},\ \Eprint {http://arxiv.org/abs/1412.5148} {arXiv:1412.5148
  [hep-th]} \BibitemShut {NoStop}%
\bibitem [{\citenamefont {Nussinov}\ and\ \citenamefont
  {Ortiz}(2009{\natexlab{a}})}]{Nussinov:2006iva}%
  \BibitemOpen
  \bibfield  {author} {\bibinfo {author} {\bibfnamefont {Z.}~\bibnamefont
  {Nussinov}}\ and\ \bibinfo {author} {\bibfnamefont {G.}~\bibnamefont
  {Ortiz}},\ }\href {\doibase 10.1073/pnas.0803726105} {\bibfield  {journal}
  {\bibinfo  {journal} {Proc. Nat. Acad. Sci.}\ }\textbf {\bibinfo {volume}
  {106}},\ \bibinfo {pages} {16944} (\bibinfo {year} {2009}{\natexlab{a}})},\
  \Eprint {http://arxiv.org/abs/cond-mat/0605316} {arXiv:cond-mat/0605316}
  \BibitemShut {NoStop}%
\bibitem [{\citenamefont {Batista}\ and\ \citenamefont
  {Nussinov}(2005)}]{Batista:2004sc}%
  \BibitemOpen
  \bibfield  {author} {\bibinfo {author} {\bibfnamefont {C.~D.}\ \bibnamefont
  {Batista}}\ and\ \bibinfo {author} {\bibfnamefont {Z.}~\bibnamefont
  {Nussinov}},\ }\href {\doibase 10.1103/PhysRevB.72.045137} {\bibfield
  {journal} {\bibinfo  {journal} {Phys. Rev. B}\ }\textbf {\bibinfo {volume}
  {72}},\ \bibinfo {pages} {045137} (\bibinfo {year} {2005})},\ \Eprint
  {http://arxiv.org/abs/cond-mat/0410599} {arXiv:cond-mat/0410599} \BibitemShut
  {NoStop}%
\bibitem [{\citenamefont {Nussinov}\ and\ \citenamefont
  {Ortiz}(2009{\natexlab{b}})}]{Nussinov:2009zz}%
  \BibitemOpen
  \bibfield  {author} {\bibinfo {author} {\bibfnamefont {Z.}~\bibnamefont
  {Nussinov}}\ and\ \bibinfo {author} {\bibfnamefont {G.}~\bibnamefont
  {Ortiz}},\ }\href {\doibase 10.1016/j.aop.2008.11.002} {\bibfield  {journal}
  {\bibinfo  {journal} {Annals Phys.}\ }\textbf {\bibinfo {volume} {324}},\
  \bibinfo {pages} {977} (\bibinfo {year} {2009}{\natexlab{b}})},\ \Eprint
  {http://arxiv.org/abs/cond-mat/0702377} {arXiv:cond-mat/0702377} \BibitemShut
  {NoStop}%
\bibitem [{\citenamefont {Nussinov}\ \emph {et~al.}(2012)\citenamefont
  {Nussinov}, \citenamefont {Ortiz},\ and\ \citenamefont
  {Cobanera}}]{Nussinov:2011mz}%
  \BibitemOpen
  \bibfield  {author} {\bibinfo {author} {\bibfnamefont {Z.}~\bibnamefont
  {Nussinov}}, \bibinfo {author} {\bibfnamefont {G.}~\bibnamefont {Ortiz}}, \
  and\ \bibinfo {author} {\bibfnamefont {E.}~\bibnamefont {Cobanera}},\ }\href
  {\doibase 10.1016/j.aop.2012.07.001} {\bibfield  {journal} {\bibinfo
  {journal} {Annals Phys.}\ }\textbf {\bibinfo {volume} {327}},\ \bibinfo
  {pages} {2491} (\bibinfo {year} {2012})},\ \Eprint
  {http://arxiv.org/abs/1110.2179} {arXiv:1110.2179 [cond-mat.stat-mech]}
  \BibitemShut {NoStop}%
\bibitem [{\citenamefont {You}\ \emph {et~al.}(2018{\natexlab{b}})\citenamefont
  {You}, \citenamefont {Devakul}, \citenamefont {Burnell},\ and\ \citenamefont
  {Sondhi}}]{PhysRevB.98.035112}%
  \BibitemOpen
  \bibfield  {author} {\bibinfo {author} {\bibfnamefont {Y.}~\bibnamefont
  {You}}, \bibinfo {author} {\bibfnamefont {T.}~\bibnamefont {Devakul}},
  \bibinfo {author} {\bibfnamefont {F.~J.}\ \bibnamefont {Burnell}}, \ and\
  \bibinfo {author} {\bibfnamefont {S.~L.}\ \bibnamefont {Sondhi}},\ }\href
  {\doibase 10.1103/PhysRevB.98.035112} {\bibfield  {journal} {\bibinfo
  {journal} {Phys. Rev. B}\ }\textbf {\bibinfo {volume} {98}},\ \bibinfo
  {pages} {035112} (\bibinfo {year} {2018}{\natexlab{b}})}\BibitemShut
  {NoStop}%
\bibitem [{\citenamefont {Devakul}\ \emph {et~al.}(2018)\citenamefont
  {Devakul}, \citenamefont {Williamson},\ and\ \citenamefont
  {You}}]{PhysRevB.98.235121}%
  \BibitemOpen
  \bibfield  {author} {\bibinfo {author} {\bibfnamefont {T.}~\bibnamefont
  {Devakul}}, \bibinfo {author} {\bibfnamefont {D.~J.}\ \bibnamefont
  {Williamson}}, \ and\ \bibinfo {author} {\bibfnamefont {Y.}~\bibnamefont
  {You}},\ }\href {\doibase 10.1103/PhysRevB.98.235121} {\bibfield  {journal}
  {\bibinfo  {journal} {Phys. Rev. B}\ }\textbf {\bibinfo {volume} {98}},\
  \bibinfo {pages} {235121} (\bibinfo {year} {2018})}\BibitemShut {NoStop}%
\bibitem [{\citenamefont {Devakul}\ \emph {et~al.}(2020)\citenamefont
  {Devakul}, \citenamefont {Shirley},\ and\ \citenamefont
  {Wang}}]{PhysRevResearch.2.012059}%
  \BibitemOpen
  \bibfield  {author} {\bibinfo {author} {\bibfnamefont {T.}~\bibnamefont
  {Devakul}}, \bibinfo {author} {\bibfnamefont {W.}~\bibnamefont {Shirley}}, \
  and\ \bibinfo {author} {\bibfnamefont {J.}~\bibnamefont {Wang}},\ }\href
  {\doibase 10.1103/PhysRevResearch.2.012059} {\bibfield  {journal} {\bibinfo
  {journal} {Phys. Rev. Research}\ }\textbf {\bibinfo {volume} {2}},\ \bibinfo
  {pages} {012059} (\bibinfo {year} {2020})}\BibitemShut {NoStop}%
\bibitem [{\citenamefont {Shen}\ \emph {et~al.}(2022)\citenamefont {Shen},
  \citenamefont {Wu}, \citenamefont {Li}, \citenamefont {Qin},\ and\
  \citenamefont {Yao}}]{PhysRevResearch.4.L032008}%
  \BibitemOpen
  \bibfield  {author} {\bibinfo {author} {\bibfnamefont {X.}~\bibnamefont
  {Shen}}, \bibinfo {author} {\bibfnamefont {Z.}~\bibnamefont {Wu}}, \bibinfo
  {author} {\bibfnamefont {L.}~\bibnamefont {Li}}, \bibinfo {author}
  {\bibfnamefont {Z.}~\bibnamefont {Qin}}, \ and\ \bibinfo {author}
  {\bibfnamefont {H.}~\bibnamefont {Yao}},\ }\href {\doibase
  10.1103/PhysRevResearch.4.L032008} {\bibfield  {journal} {\bibinfo  {journal}
  {Phys. Rev. Research}\ }\textbf {\bibinfo {volume} {4}},\ \bibinfo {pages}
  {L032008} (\bibinfo {year} {2022})}\BibitemShut {NoStop}%
\bibitem [{\citenamefont {Tsui}\ \emph {et~al.}(2015)\citenamefont {Tsui},
  \citenamefont {Jiang}, \citenamefont {Lu},\ and\ \citenamefont
  {Lee}}]{tsui2015quantum}%
  \BibitemOpen
  \bibfield  {author} {\bibinfo {author} {\bibfnamefont {L.}~\bibnamefont
  {Tsui}}, \bibinfo {author} {\bibfnamefont {H.-C.}\ \bibnamefont {Jiang}},
  \bibinfo {author} {\bibfnamefont {Y.-M.}\ \bibnamefont {Lu}}, \ and\ \bibinfo
  {author} {\bibfnamefont {D.-H.}\ \bibnamefont {Lee}},\ }\href@noop {}
  {\bibfield  {journal} {\bibinfo  {journal} {Nuclear Physics B}\ }\textbf
  {\bibinfo {volume} {896}},\ \bibinfo {pages} {330} (\bibinfo {year}
  {2015})}\BibitemShut {NoStop}%
\bibitem [{\citenamefont {Bultinck}(2019)}]{PhysRevB.100.165132}%
  \BibitemOpen
  \bibfield  {author} {\bibinfo {author} {\bibfnamefont {N.}~\bibnamefont
  {Bultinck}},\ }\href {\doibase 10.1103/PhysRevB.100.165132} {\bibfield
  {journal} {\bibinfo  {journal} {Phys. Rev. B}\ }\textbf {\bibinfo {volume}
  {100}},\ \bibinfo {pages} {165132} (\bibinfo {year} {2019})}\BibitemShut
  {NoStop}%
\bibitem [{\citenamefont {'t~Hooft}\ \emph {et~al.}(1980)\citenamefont
  {'t~Hooft}, \citenamefont {Itzykson}, \citenamefont {Jaffe}, \citenamefont
  {Lehmann}, \citenamefont {Mitter}, \citenamefont {Singer},\ and\
  \citenamefont {Stora}}]{tHooft:1980xss}%
  \BibitemOpen
  \bibinfo {editor} {\bibfnamefont {G.}~\bibnamefont {'t~Hooft}}, \bibinfo
  {editor} {\bibfnamefont {C.}~\bibnamefont {Itzykson}}, \bibinfo {editor}
  {\bibfnamefont {A.}~\bibnamefont {Jaffe}}, \bibinfo {editor} {\bibfnamefont
  {H.}~\bibnamefont {Lehmann}}, \bibinfo {editor} {\bibfnamefont {P.~K.}\
  \bibnamefont {Mitter}}, \bibinfo {editor} {\bibfnamefont {I.~M.}\
  \bibnamefont {Singer}}, \ and\ \bibinfo {editor} {\bibfnamefont
  {R.}~\bibnamefont {Stora}},\ eds.,\ \href {\doibase
  10.1007/978-1-4684-7571-5} {\emph {\bibinfo {title} {{Recent Developments in
  Gauge Theories. Proceedings, Nato Advanced Study Institute, Cargese, France,
  August 26 - September 8, 1979}}}},\ Vol.~\bibinfo {volume} {59}\ (\bibinfo
  {year} {1980})\BibitemShut {NoStop}%
\bibitem [{\citenamefont {Lieb}\ \emph {et~al.}(1961)\citenamefont {Lieb},
  \citenamefont {Schultz},\ and\ \citenamefont {Mattis}}]{lieb1961two}%
  \BibitemOpen
  \bibfield  {author} {\bibinfo {author} {\bibfnamefont {E.}~\bibnamefont
  {Lieb}}, \bibinfo {author} {\bibfnamefont {T.}~\bibnamefont {Schultz}}, \
  and\ \bibinfo {author} {\bibfnamefont {D.}~\bibnamefont {Mattis}},\
  }\href@noop {} {\bibfield  {journal} {\bibinfo  {journal} {Annals of
  Physics}\ }\textbf {\bibinfo {volume} {16}},\ \bibinfo {pages} {407}
  (\bibinfo {year} {1961})}\BibitemShut {NoStop}%
\bibitem [{\citenamefont {Affleck}\ and\ \citenamefont
  {Lieb}(2004)}]{Affleck2004}%
  \BibitemOpen
  \bibfield  {author} {\bibinfo {author} {\bibfnamefont {I.}~\bibnamefont
  {Affleck}}\ and\ \bibinfo {author} {\bibfnamefont {E.~H.}\ \bibnamefont
  {Lieb}},\ }\enquote {\bibinfo {title} {A proof of part of haldane's
  conjecture on spin chains},}\ in\ \href {\doibase
  10.1007/978-3-662-06390-3_17} {\emph {\bibinfo {booktitle} {Condensed Matter
  Physics and Exactly Soluble Models: Selecta of Elliott H. Lieb}}},\ \bibinfo
  {editor} {edited by\ \bibinfo {editor} {\bibfnamefont {B.}~\bibnamefont
  {Nachtergaele}}, \bibinfo {editor} {\bibfnamefont {J.~P.}\ \bibnamefont
  {Solovej}}, \ and\ \bibinfo {editor} {\bibfnamefont {J.}~\bibnamefont
  {Yngvason}}}\ (\bibinfo  {publisher} {Springer Berlin Heidelberg},\ \bibinfo
  {address} {Berlin, Heidelberg},\ \bibinfo {year} {2004})\ pp.\ \bibinfo
  {pages} {235--247}\BibitemShut {NoStop}%
\bibitem [{\citenamefont {Oshikawa}\ \emph {et~al.}(1997)\citenamefont
  {Oshikawa}, \citenamefont {Yamanaka},\ and\ \citenamefont
  {Affleck}}]{PhysRevLett.78.1984}%
  \BibitemOpen
  \bibfield  {author} {\bibinfo {author} {\bibfnamefont {M.}~\bibnamefont
  {Oshikawa}}, \bibinfo {author} {\bibfnamefont {M.}~\bibnamefont {Yamanaka}},
  \ and\ \bibinfo {author} {\bibfnamefont {I.}~\bibnamefont {Affleck}},\ }\href
  {\doibase 10.1103/PhysRevLett.78.1984} {\bibfield  {journal} {\bibinfo
  {journal} {Phys. Rev. Lett.}\ }\textbf {\bibinfo {volume} {78}},\ \bibinfo
  {pages} {1984} (\bibinfo {year} {1997})}\BibitemShut {NoStop}%
\bibitem [{\citenamefont {Oshikawa}(2000)}]{PhysRevLett.84.1535}%
  \BibitemOpen
  \bibfield  {author} {\bibinfo {author} {\bibfnamefont {M.}~\bibnamefont
  {Oshikawa}},\ }\href {\doibase 10.1103/PhysRevLett.84.1535} {\bibfield
  {journal} {\bibinfo  {journal} {Phys. Rev. Lett.}\ }\textbf {\bibinfo
  {volume} {84}},\ \bibinfo {pages} {1535} (\bibinfo {year}
  {2000})}\BibitemShut {NoStop}%
\bibitem [{\citenamefont {Hastings}(2004)}]{PhysRevB.69.104431}%
  \BibitemOpen
  \bibfield  {author} {\bibinfo {author} {\bibfnamefont {M.~B.}\ \bibnamefont
  {Hastings}},\ }\href {\doibase 10.1103/PhysRevB.69.104431} {\bibfield
  {journal} {\bibinfo  {journal} {Phys. Rev. B}\ }\textbf {\bibinfo {volume}
  {69}},\ \bibinfo {pages} {104431} (\bibinfo {year} {2004})}\BibitemShut
  {NoStop}%
\bibitem [{\citenamefont {Furuya}\ and\ \citenamefont
  {Oshikawa}(2017)}]{PhysRevLett.118.021601}%
  \BibitemOpen
  \bibfield  {author} {\bibinfo {author} {\bibfnamefont {S.~C.}\ \bibnamefont
  {Furuya}}\ and\ \bibinfo {author} {\bibfnamefont {M.}~\bibnamefont
  {Oshikawa}},\ }\href {\doibase 10.1103/PhysRevLett.118.021601} {\bibfield
  {journal} {\bibinfo  {journal} {Phys. Rev. Lett.}\ }\textbf {\bibinfo
  {volume} {118}},\ \bibinfo {pages} {021601} (\bibinfo {year}
  {2017})}\BibitemShut {NoStop}%
\bibitem [{\citenamefont {Yao}\ \emph {et~al.}(2019)\citenamefont {Yao},
  \citenamefont {Hsieh},\ and\ \citenamefont {Oshikawa}}]{yao2019anomaly}%
  \BibitemOpen
  \bibfield  {author} {\bibinfo {author} {\bibfnamefont {Y.}~\bibnamefont
  {Yao}}, \bibinfo {author} {\bibfnamefont {C.-T.}\ \bibnamefont {Hsieh}}, \
  and\ \bibinfo {author} {\bibfnamefont {M.}~\bibnamefont {Oshikawa}},\
  }\href@noop {} {\bibfield  {journal} {\bibinfo  {journal} {Physical review
  letters}\ }\textbf {\bibinfo {volume} {123}},\ \bibinfo {pages} {180201}
  (\bibinfo {year} {2019})}\BibitemShut {NoStop}%
\bibitem [{\citenamefont {Aksoy}\ \emph
  {et~al.}(2021{\natexlab{b}})\citenamefont {Aksoy}, \citenamefont {Tiwari},\
  and\ \citenamefont {Mudry}}]{aksoy2021lieb}%
  \BibitemOpen
  \bibfield  {author} {\bibinfo {author} {\bibfnamefont {{\"O}.~M.}\
  \bibnamefont {Aksoy}}, \bibinfo {author} {\bibfnamefont {A.}~\bibnamefont
  {Tiwari}}, \ and\ \bibinfo {author} {\bibfnamefont {C.}~\bibnamefont
  {Mudry}},\ }\href@noop {} {\bibfield  {journal} {\bibinfo  {journal}
  {Physical Review B}\ }\textbf {\bibinfo {volume} {104}},\ \bibinfo {pages}
  {075146} (\bibinfo {year} {2021}{\natexlab{b}})}\BibitemShut {NoStop}%
\bibitem [{\citenamefont {Li}\ \emph {et~al.}(2022{\natexlab{a}})\citenamefont
  {Li}, \citenamefont {Hsieh}, \citenamefont {Yao},\ and\ \citenamefont
  {Oshikawa}}]{li2022boundary}%
  \BibitemOpen
  \bibfield  {author} {\bibinfo {author} {\bibfnamefont {L.}~\bibnamefont
  {Li}}, \bibinfo {author} {\bibfnamefont {C.-T.}\ \bibnamefont {Hsieh}},
  \bibinfo {author} {\bibfnamefont {Y.}~\bibnamefont {Yao}}, \ and\ \bibinfo
  {author} {\bibfnamefont {M.}~\bibnamefont {Oshikawa}},\ }\href@noop {}
  {\bibfield  {journal} {\bibinfo  {journal} {arXiv preprint arXiv:2205.11190}\
  } (\bibinfo {year} {2022}{\natexlab{a}})}\BibitemShut {NoStop}%
\bibitem [{\citenamefont {Watanabe}(2018)}]{PhysRevB.98.155137}%
  \BibitemOpen
  \bibfield  {author} {\bibinfo {author} {\bibfnamefont {H.}~\bibnamefont
  {Watanabe}},\ }\href {\doibase 10.1103/PhysRevB.98.155137} {\bibfield
  {journal} {\bibinfo  {journal} {Phys. Rev. B}\ }\textbf {\bibinfo {volume}
  {98}},\ \bibinfo {pages} {155137} (\bibinfo {year} {2018})}\BibitemShut
  {NoStop}%
\bibitem [{\citenamefont {Yao}\ and\ \citenamefont
  {Oshikawa}(2021)}]{yao2021twisted}%
  \BibitemOpen
  \bibfield  {author} {\bibinfo {author} {\bibfnamefont {Y.}~\bibnamefont
  {Yao}}\ and\ \bibinfo {author} {\bibfnamefont {M.}~\bibnamefont {Oshikawa}},\
  }\href@noop {} {\bibfield  {journal} {\bibinfo  {journal} {Physical Review
  Letters}\ }\textbf {\bibinfo {volume} {126}},\ \bibinfo {pages} {217201}
  (\bibinfo {year} {2021})}\BibitemShut {NoStop}%
\bibitem [{\citenamefont {Yao}\ and\ \citenamefont
  {Furusaki}(2022)}]{yao2022geometric}%
  \BibitemOpen
  \bibfield  {author} {\bibinfo {author} {\bibfnamefont {Y.}~\bibnamefont
  {Yao}}\ and\ \bibinfo {author} {\bibfnamefont {A.}~\bibnamefont {Furusaki}},\
  }\href@noop {} {\bibfield  {journal} {\bibinfo  {journal} {Physical Review
  B}\ }\textbf {\bibinfo {volume} {106}},\ \bibinfo {pages} {045125} (\bibinfo
  {year} {2022})}\BibitemShut {NoStop}%
\bibitem [{\citenamefont {Scaffidi}\ \emph {et~al.}(2017)\citenamefont
  {Scaffidi}, \citenamefont {Parker},\ and\ \citenamefont
  {Vasseur}}]{scaffidi2017gapless}%
  \BibitemOpen
  \bibfield  {author} {\bibinfo {author} {\bibfnamefont {T.}~\bibnamefont
  {Scaffidi}}, \bibinfo {author} {\bibfnamefont {D.~E.}\ \bibnamefont
  {Parker}}, \ and\ \bibinfo {author} {\bibfnamefont {R.}~\bibnamefont
  {Vasseur}},\ }\href@noop {} {\bibfield  {journal} {\bibinfo  {journal}
  {Physical Review X}\ }\textbf {\bibinfo {volume} {7}},\ \bibinfo {pages}
  {041048} (\bibinfo {year} {2017})}\BibitemShut {NoStop}%
\bibitem [{\citenamefont {Parker}\ \emph {et~al.}(2018)\citenamefont {Parker},
  \citenamefont {Scaffidi},\ and\ \citenamefont
  {Vasseur}}]{PhysRevB.97.165114}%
  \BibitemOpen
  \bibfield  {author} {\bibinfo {author} {\bibfnamefont {D.~E.}\ \bibnamefont
  {Parker}}, \bibinfo {author} {\bibfnamefont {T.}~\bibnamefont {Scaffidi}}, \
  and\ \bibinfo {author} {\bibfnamefont {R.}~\bibnamefont {Vasseur}},\ }\href
  {\doibase 10.1103/PhysRevB.97.165114} {\bibfield  {journal} {\bibinfo
  {journal} {Phys. Rev. B}\ }\textbf {\bibinfo {volume} {97}},\ \bibinfo
  {pages} {165114} (\bibinfo {year} {2018})}\BibitemShut {NoStop}%
\bibitem [{\citenamefont {Parker}\ \emph {et~al.}(2019)\citenamefont {Parker},
  \citenamefont {Vasseur},\ and\ \citenamefont
  {Scaffidi}}]{PhysRevLett.122.240605}%
  \BibitemOpen
  \bibfield  {author} {\bibinfo {author} {\bibfnamefont {D.~E.}\ \bibnamefont
  {Parker}}, \bibinfo {author} {\bibfnamefont {R.}~\bibnamefont {Vasseur}}, \
  and\ \bibinfo {author} {\bibfnamefont {T.}~\bibnamefont {Scaffidi}},\ }\href
  {\doibase 10.1103/PhysRevLett.122.240605} {\bibfield  {journal} {\bibinfo
  {journal} {Phys. Rev. Lett.}\ }\textbf {\bibinfo {volume} {122}},\ \bibinfo
  {pages} {240605} (\bibinfo {year} {2019})}\BibitemShut {NoStop}%
\bibitem [{\citenamefont {Thorngren}\ \emph {et~al.}(2020)\citenamefont
  {Thorngren}, \citenamefont {Vishwanath},\ and\ \citenamefont
  {Verresen}}]{Thorngren:2020wet}%
  \BibitemOpen
  \bibfield  {author} {\bibinfo {author} {\bibfnamefont {R.}~\bibnamefont
  {Thorngren}}, \bibinfo {author} {\bibfnamefont {A.}~\bibnamefont
  {Vishwanath}}, \ and\ \bibinfo {author} {\bibfnamefont {R.}~\bibnamefont
  {Verresen}},\ }\href@noop {} {\  (\bibinfo {year} {2020})},\ \Eprint
  {http://arxiv.org/abs/2008.06638} {arXiv:2008.06638 [cond-mat.str-el]}
  \BibitemShut {NoStop}%
\bibitem [{\citenamefont {Wang}\ \emph
  {et~al.}(2021{\natexlab{b}})\citenamefont {Wang}, \citenamefont {Ning},\ and\
  \citenamefont {Cheng}}]{Wang:2021nrp}%
  \BibitemOpen
  \bibfield  {author} {\bibinfo {author} {\bibfnamefont {Q.-R.}\ \bibnamefont
  {Wang}}, \bibinfo {author} {\bibfnamefont {S.-Q.}\ \bibnamefont {Ning}}, \
  and\ \bibinfo {author} {\bibfnamefont {M.}~\bibnamefont {Cheng}},\
  }\href@noop {} {\  (\bibinfo {year} {2021}{\natexlab{b}})},\ \Eprint
  {http://arxiv.org/abs/2104.13233} {arXiv:2104.13233 [cond-mat.str-el]}
  \BibitemShut {NoStop}%
\bibitem [{\citenamefont {Li}\ \emph {et~al.}(2022{\natexlab{b}})\citenamefont
  {Li}, \citenamefont {Oshikawa},\ and\ \citenamefont
  {Zheng}}]{li2022symmetry}%
  \BibitemOpen
  \bibfield  {author} {\bibinfo {author} {\bibfnamefont {L.}~\bibnamefont
  {Li}}, \bibinfo {author} {\bibfnamefont {M.}~\bibnamefont {Oshikawa}}, \ and\
  \bibinfo {author} {\bibfnamefont {Y.}~\bibnamefont {Zheng}},\ }\href@noop {}
  {\bibfield  {journal} {\bibinfo  {journal} {arXiv preprint arXiv:2204.03131}\
  } (\bibinfo {year} {2022}{\natexlab{b}})}\BibitemShut {NoStop}%
\bibitem [{\citenamefont {Xu}\ and\ \citenamefont
  {Moore}(2004)}]{PhysRevLett.93.047003}%
  \BibitemOpen
  \bibfield  {author} {\bibinfo {author} {\bibfnamefont {C.}~\bibnamefont
  {Xu}}\ and\ \bibinfo {author} {\bibfnamefont {J.~E.}\ \bibnamefont {Moore}},\
  }\href {\doibase 10.1103/PhysRevLett.93.047003} {\bibfield  {journal}
  {\bibinfo  {journal} {Phys. Rev. Lett.}\ }\textbf {\bibinfo {volume} {93}},\
  \bibinfo {pages} {047003} (\bibinfo {year} {2004})}\BibitemShut {NoStop}%
\bibitem [{\citenamefont {Or\'us}\ \emph {et~al.}(2009)\citenamefont {Or\'us},
  \citenamefont {Doherty},\ and\ \citenamefont
  {Vidal}}]{PhysRevLett.102.077203}%
  \BibitemOpen
  \bibfield  {author} {\bibinfo {author} {\bibfnamefont {R.}~\bibnamefont
  {Or\'us}}, \bibinfo {author} {\bibfnamefont {A.~C.}\ \bibnamefont {Doherty}},
  \ and\ \bibinfo {author} {\bibfnamefont {G.}~\bibnamefont {Vidal}},\ }\href
  {\doibase 10.1103/PhysRevLett.102.077203} {\bibfield  {journal} {\bibinfo
  {journal} {Phys. Rev. Lett.}\ }\textbf {\bibinfo {volume} {102}},\ \bibinfo
  {pages} {077203} (\bibinfo {year} {2009})}\BibitemShut {NoStop}%
\bibitem [{\citenamefont {Kalis}\ \emph {et~al.}(2012)\citenamefont {Kalis},
  \citenamefont {Klagges}, \citenamefont {Or\'us},\ and\ \citenamefont
  {Schmidt}}]{PhysRevA.86.022317}%
  \BibitemOpen
  \bibfield  {author} {\bibinfo {author} {\bibfnamefont {H.}~\bibnamefont
  {Kalis}}, \bibinfo {author} {\bibfnamefont {D.}~\bibnamefont {Klagges}},
  \bibinfo {author} {\bibfnamefont {R.}~\bibnamefont {Or\'us}}, \ and\ \bibinfo
  {author} {\bibfnamefont {K.~P.}\ \bibnamefont {Schmidt}},\ }\href {\doibase
  10.1103/PhysRevA.86.022317} {\bibfield  {journal} {\bibinfo  {journal} {Phys.
  Rev. A}\ }\textbf {\bibinfo {volume} {86}},\ \bibinfo {pages} {022317}
  (\bibinfo {year} {2012})}\BibitemShut {NoStop}%
\bibitem [{\citenamefont {Zhou}\ \emph {et~al.}(2022)\citenamefont {Zhou},
  \citenamefont {Li}, \citenamefont {Yan}, \citenamefont {Ye},\ and\
  \citenamefont {Meng}}]{Zhou:2022eig}%
  \BibitemOpen
  \bibfield  {author} {\bibinfo {author} {\bibfnamefont {C.}~\bibnamefont
  {Zhou}}, \bibinfo {author} {\bibfnamefont {M.-Y.}\ \bibnamefont {Li}},
  \bibinfo {author} {\bibfnamefont {Z.}~\bibnamefont {Yan}}, \bibinfo {author}
  {\bibfnamefont {P.}~\bibnamefont {Ye}}, \ and\ \bibinfo {author}
  {\bibfnamefont {Z.~Y.}\ \bibnamefont {Meng}},\ }\href@noop {} {\  (\bibinfo
  {year} {2022})},\ \Eprint {http://arxiv.org/abs/2209.12917} {arXiv:2209.12917
  [cond-mat.str-el]} \BibitemShut {NoStop}%
\bibitem [{\citenamefont {Tantivasadakarn}\ \emph
  {et~al.}(2021{\natexlab{b}})\citenamefont {Tantivasadakarn}, \citenamefont
  {Thorngren}, \citenamefont {Vishwanath},\ and\ \citenamefont
  {Verresen}}]{Tantivasadakarn:2021wdv}%
  \BibitemOpen
  \bibfield  {author} {\bibinfo {author} {\bibfnamefont {N.}~\bibnamefont
  {Tantivasadakarn}}, \bibinfo {author} {\bibfnamefont {R.}~\bibnamefont
  {Thorngren}}, \bibinfo {author} {\bibfnamefont {A.}~\bibnamefont
  {Vishwanath}}, \ and\ \bibinfo {author} {\bibfnamefont {R.}~\bibnamefont
  {Verresen}},\ }\href@noop {} {\  (\bibinfo {year} {2021}{\natexlab{b}})},\
  \Eprint {http://arxiv.org/abs/2110.07599} {arXiv:2110.07599
  [cond-mat.str-el]} \BibitemShut {NoStop}%
\bibitem [{\citenamefont {Suzuki}(1971)}]{suzuki1971relationship}%
  \BibitemOpen
  \bibfield  {author} {\bibinfo {author} {\bibfnamefont {M.}~\bibnamefont
  {Suzuki}},\ }\href@noop {} {\bibfield  {journal} {\bibinfo  {journal}
  {Progress of Theoretical Physics}\ }\textbf {\bibinfo {volume} {46}},\
  \bibinfo {pages} {1337} (\bibinfo {year} {1971})}\BibitemShut {NoStop}%
\bibitem [{\citenamefont {Keating}\ and\ \citenamefont
  {Mezzadri}(2004)}]{keating2004random}%
  \BibitemOpen
  \bibfield  {author} {\bibinfo {author} {\bibfnamefont {J.~P.}\ \bibnamefont
  {Keating}}\ and\ \bibinfo {author} {\bibfnamefont {F.}~\bibnamefont
  {Mezzadri}},\ }\href@noop {} {\bibfield  {journal} {\bibinfo  {journal}
  {Communications in mathematical physics}\ }\textbf {\bibinfo {volume}
  {252}},\ \bibinfo {pages} {543} (\bibinfo {year} {2004})}\BibitemShut
  {NoStop}%
\bibitem [{\citenamefont {Metlitski}\ and\ \citenamefont
  {Thorngren}(2018)}]{metlitski2018intrinsic}%
  \BibitemOpen
  \bibfield  {author} {\bibinfo {author} {\bibfnamefont {M.~A.}\ \bibnamefont
  {Metlitski}}\ and\ \bibinfo {author} {\bibfnamefont {R.}~\bibnamefont
  {Thorngren}},\ }\href@noop {} {\bibfield  {journal} {\bibinfo  {journal}
  {Physical Review B}\ }\textbf {\bibinfo {volume} {98}},\ \bibinfo {pages}
  {085140} (\bibinfo {year} {2018})}\BibitemShut {NoStop}%
\bibitem [{Note1()}]{Note1}%
  \BibitemOpen
  \bibinfo {note} {Here we treat $\tau $ and $\sigma $ as one kind of spin in
  Kramers-Wannier transformation.}\BibitemShut {Stop}%
\bibitem [{Note2()}]{Note2}%
  \BibitemOpen
  \bibinfo {note} {If we include the time-reversal symmetry $T=K$ which is
  complex conjugation, these two terms and the cluster chain \protect \eqref
  {cluster} are distinct nontrivial SPT phases.}\BibitemShut {Stop}%
\bibitem [{Note3()}]{Note3}%
  \BibitemOpen
  \bibinfo {note} {If the length $L$ is not a multiple of 3, this Hamiltonian
  is equivalent to the nearest neighboured hopping term with a long rang
  interaction when $J\protect \ne 0$}\BibitemShut {NoStop}%
\bibitem [{Note4()}]{Note4}%
  \BibitemOpen
  \bibinfo {note} {The configuration with lowest energy is $(\varphi _1-\varphi
  _{2},\varphi _2-\varphi _{3})=(\protect \frac {\pi }{3},\protect \frac {\pi
  }{3})/(-\protect \frac {\pi }{3},-\protect \frac {\pi }{3})$}\BibitemShut
  {NoStop}%
\bibitem [{\citenamefont {Dupont}\ \emph
  {et~al.}(2021{\natexlab{a}})\citenamefont {Dupont}, \citenamefont {Gazit},\
  and\ \citenamefont {Scaffidi}}]{PhysRevB.103.L140412}%
  \BibitemOpen
  \bibfield  {author} {\bibinfo {author} {\bibfnamefont {M.}~\bibnamefont
  {Dupont}}, \bibinfo {author} {\bibfnamefont {S.}~\bibnamefont {Gazit}}, \
  and\ \bibinfo {author} {\bibfnamefont {T.}~\bibnamefont {Scaffidi}},\ }\href
  {\doibase 10.1103/PhysRevB.103.L140412} {\bibfield  {journal} {\bibinfo
  {journal} {Phys. Rev. B}\ }\textbf {\bibinfo {volume} {103}},\ \bibinfo
  {pages} {L140412} (\bibinfo {year} {2021}{\natexlab{a}})}\BibitemShut
  {NoStop}%
\bibitem [{\citenamefont {Dupont}\ \emph
  {et~al.}(2021{\natexlab{b}})\citenamefont {Dupont}, \citenamefont {Gazit},\
  and\ \citenamefont {Scaffidi}}]{PhysRevB.103.144437}%
  \BibitemOpen
  \bibfield  {author} {\bibinfo {author} {\bibfnamefont {M.}~\bibnamefont
  {Dupont}}, \bibinfo {author} {\bibfnamefont {S.}~\bibnamefont {Gazit}}, \
  and\ \bibinfo {author} {\bibfnamefont {T.}~\bibnamefont {Scaffidi}},\ }\href
  {\doibase 10.1103/PhysRevB.103.144437} {\bibfield  {journal} {\bibinfo
  {journal} {Phys. Rev. B}\ }\textbf {\bibinfo {volume} {103}},\ \bibinfo
  {pages} {144437} (\bibinfo {year} {2021}{\natexlab{b}})}\BibitemShut
  {NoStop}%
\bibitem [{\citenamefont {Hsieh}\ \emph {et~al.}(2016)\citenamefont {Hsieh},
  \citenamefont {Hal\'asz},\ and\ \citenamefont
  {Grover}}]{PhysRevLett.117.166802}%
  \BibitemOpen
  \bibfield  {author} {\bibinfo {author} {\bibfnamefont {T.~H.}\ \bibnamefont
  {Hsieh}}, \bibinfo {author} {\bibfnamefont {G.~B.}\ \bibnamefont {Hal\'asz}},
  \ and\ \bibinfo {author} {\bibfnamefont {T.}~\bibnamefont {Grover}},\ }\href
  {\doibase 10.1103/PhysRevLett.117.166802} {\bibfield  {journal} {\bibinfo
  {journal} {Phys. Rev. Lett.}\ }\textbf {\bibinfo {volume} {117}},\ \bibinfo
  {pages} {166802} (\bibinfo {year} {2016})}\BibitemShut {NoStop}%
\end{thebibliography}%
	\clearpage
	
	\appendix
	\begin{widetext}
	\section{Ingappabilities of 1+1 dimensional models with $\mathbb{Z}^{(0,0)}_N$ and $\mathbb{Z}_N\times \mathbb{Z}_N $ symmetry }\label{SM1}
In this appendix, we generalize the result in section \ref{1dproof}  to the $\mathbb{Z}_N\times \mathbb{Z}_N $ symmetry. Let us place two $\mathbb{Z}_{N}$ "spins" $\sigma$ and $\tau$ on the integer and half-integer sites respectively. They satisfy the $\mathbb{Z}_N$ Heisenberg algebra:
\begin{eqnarray}\label{Heisenberg algebra}
(\sigma^x)^{N}=(\sigma^z)^{N}=1,\quad \sigma^z\sigma^x=\omega\sigma^x\sigma^z,\quad (\sigma^z)^{\dagger}=(\sigma^z)^{N-1},\quad (\sigma^x)^{\dagger}=(\sigma^x)^{N-1},\\
(\tau^x)^{N}=(\tau^z)^{N}=1,\quad \tau^z\tau^x=\omega\tau^x\tau^z,\quad (\tau^z)^{\dagger}=(\tau^z)^{N-1},\quad (\tau^x)^{\dagger}=(\tau^x)^{N-1},
\end{eqnarray}
where $\omega=\exp(\frac{2\pi i}{N}).$

The natural generalization of the paramagnetic Hamiltonian to the $N$-state case is 
\begin{eqnarray}\label{ZNssb}
H_{0}=-\sum^L_{j=1} (\sigma^x_j+\tau^x_{j+\frac{1}{2}}+h.c.)
\end{eqnarray}
where we assume the periodic boundary condition:
$\sigma^{x/z}_{i+L}=\sigma^{x/z}_{i}$ and $\tau^{x/z}_{i+L}=\tau^{x/z}_{i}$.

This Hamiltonian has $\mathbb{Z}^A_N\times \mathbb{Z}^G_N$ symmetry:
\begin{eqnarray}
U_{G}=\prod_{j}\sigma^x_j,\quad U_{A}=\prod_{j}\tau^x_{j+\frac{1}{2}}.
\end{eqnarray}

To be transparent, let us define an $N$-dimensional basis
on each site, satisfying
\begin{eqnarray}\label{representation}
&&\sigma^z_{j}|\alpha\rangle_j=
\omega^{\alpha}|\alpha\rangle_j,\quad  \sigma^x_{j}|\alpha\rangle=
|\alpha+1  (\text{mod $N$})\rangle_j\\
&&\tau^z_{j+\frac{1}{2}}|\beta\rangle_{j+\frac{1}{2}}=
\omega^{\beta}|\beta\rangle_{j+\frac{1}{2}},\quad  \tau^x_{j+\frac{1}{2}}|\beta\rangle_{j+\frac{1}{2}}=
|\beta+1  (\text{mod $N$})\rangle_{j+\frac{1}{2}}
\end{eqnarray}
for $\alpha$ $\in$ $\mathbb{Z}_N$ and $\beta$ $\in$ $\mathbb{Z}_N$.

Moreover, we can define a duality operator  $U_{(0,0)}$ as follows:
\begin{eqnarray}
&&U_{(0,0)}=\prod^{L}_{i=1} \text{CZ}_{i,i+\frac{1}{2}}\text{CZ}^{-1}_{i+\frac{1}{2},i+1},\nonumber\\ &&\text{CZ}_{i,i+\frac{1}{2}}|\alpha,\beta\rangle_{i,i+\frac{1}{2}}=\omega^{\alpha\beta}|\alpha,\beta\rangle_{i,i+\frac{1}{2}},\quad \text{CZ}^{-1}_{i+\frac{1}{2},i+1}|\alpha,\beta\rangle_{i+\frac{1}{2},i+1}=\omega^{-\alpha\beta}|\alpha,\beta\rangle_{i+\frac{1}{2},i+1}.
\end{eqnarray}
After conjugating the paramagnetic Hamiltonian by this transformation, we obtain
\begin{eqnarray}\label{ZNSPT}
H_{1}:=U_{(0,0)}H_0 U^{-1}_{(0,0)}=-\sum^L_{j=1} [(\tau^z_{j-\frac{1}{2}})^{-1}\sigma^x_j \tau^z_{j+\frac{1}{2}}+\sigma^z_j\tau^x_{j+\frac{1}{2}}(\sigma^z_{j+1})^{-1}],
\end{eqnarray}
which is $\mathbb{Z}_N$ cluster model and corresponds to the generator of the group cohomology $H^2(\mathbb{Z}_N\times \mathbb{Z}_N, U(1))=\mathbb{Z}_N $. Moreover,  we can continue to conjugate \eqref{ZNSPT} by this transformation and obtain all SPT phases:
\begin{eqnarray}
H_{k}:=U^k_{(0,0)}H_0 U^{-k}_{(0,0)}=-\sum^L_{j=1} [(\tau^z_{j-\frac{1}{2}})^{-k}\sigma^x_j (\tau^z_{j+\frac{1}{2}})^k+(\sigma^z_j)^k\tau^x_{j+\frac{1}{2}}(\sigma^z_{j+1})^{-k}].
\end{eqnarray}
Since $U_{(0,0)}$ is a $\mathbb{Z}_N$ transformation, $k\in \mathbb{Z}_N $. Then the self-dual model can be a combination of all SPT phases:
\begin{eqnarray}
H_{\text{self-dual}}=\sum^{N-1}_{k=0} H_k.
\end{eqnarray}

To prove ingappabilities of the general self-dual models, we can consider symmetry twisted boundary condition using the $\mathbb{Z}_{N}^A$ symmetry:
\begin{eqnarray}
\tau_{i+L-\frac{1}{2}}^{a}=U^{-1}_A \tau_{i-\frac{1}{2}}^{a}U_A, \quad i=1,\cdots,L,
\end{eqnarray}
where the closed boundary bond is between the sites $i=L$ and $i=1$.

Following the same calculation in the section \ref{1dproof}, the modified duality transformation is given by
\begin{eqnarray}
U^{(1)}_{(0,0)}=\sigma^z_L U_{(0,0)},
\end{eqnarray}
which has a nontrivial commutator with the $U_G$ operator:
\begin{eqnarray}
U^{(1)}_{(0,0)}U_G=\exp({\frac{2\pi i}{N}}) U_G U^{(1)}_{(0,0)}.
\end{eqnarray}

 Hence, the twisted Hamiltonian must have an exactly $N$ degenerate spectrum. According to the spectrum robustness argument, the Hamiltonian under PBC must either be gapless or have a nontrivial ground-state degeneracy, rather than a unique gapped ground state.

Besides, one can also prove the ingappability if a model only preserves any nontrivial subgroup of $\mathbb{Z}^{(0,0)}_{N}$ symmetry with additional $\mathbb{Z}^A_N\times \mathbb{Z}^G_N$ symmetry. For example, let us consider a group generating by $U_{(0,0)}^m$ where $N$ divides $m$. If we twist the boundary condition using the $\mathbb{Z}_{N}^A$ symmetry, then the modified duality transformation is given by
\begin{eqnarray}
(U^{(1)}_{(0,0)})^m=(\sigma^z_L U_{(0,0)})^m
\end{eqnarray}
which still does not commute with the $U_G$ operator. 

Thus we can conclude that if a $\mathbb{Z}^A_N\times \mathbb{Z}^G_N$ invariant spin chain also possesses any nontrivial subgroup of $\mathbb{Z}^{(0,0)}_{N}$ symmetry, the Hamiltonian under PBC can be gapped only if the ground states are at least doubly degenerate.

\section{Proof of the duality operator under twisted boundary condition on the closed lattcie}\label{SM2}

 In this appendix, we will prove Eq.\eqref{duality2d} and Eq.\eqref{dualityanyd} under symmetry twisted boundary conditions on the closed lattice.

 Let us begin with a $d$ dimensional simplicial lattice with a twisted boundary condition using $U_{a_1}$ operator in  section \ref{anydproof}. As discussed before, this twisted boundary is close and locally parallel to a closed and connected $d-1$ dimensional sublattice $S_1$ which is colored in $a_2,\cdots,a_{d+1}$. When $d=2$, the twisted boundary and the corresponding one dimensional sublattice are shown in   Fig.~\ref{sigmatwist}.

 To manifest how the twisted boundary condition modifies the duality operator,  we consider a local term $H_{0}$ in the Hamiltonian with PBC and its dual term by conjugating the duality operator. Here we denote the union of the regions of $H_0$ and its dual term $H_{\text{dual}}$ as $R$.  Similar to the discussion in section \ref{1dproof}, the pair of this local term and its dual term can be classified into three cases. The first and second cases are that both two terms cross or do not cross the twisted boundary. And in the last case, we can assume only the dual term $H_{\text{dual}}$ does not cross the twisted boundary without loss of generality.

In the first case, the region $R$ is divided by the twisted boundary into the up part $R_1$ and the down part $R_2$. 
Then after twisting, we have $H^{\text{tw}}_{0}=(\prod_{v\in R'}X^{(a_1)}_v) H_0 (\prod_{v\in R'}X^{(a_1)}_v)$ and  $H^{\text{tw}}_{\text{dual}}=(\prod_{v\in R'}X^{(a_1)}_v) H_{\text{dual}} (\prod_{v\in R'}X^{(a_1)}_v)$  where the sublattice $R'$ is on the same side of $R_2$ and $R_2 \subset R'$.   Besides, we can construct this region $R'$ so that $\partial R'=S' \oplus S_1$ where $S'$ is another closed and connected $d-1$ dimensional sublattice colored in $a_2,\cdots,a_{d+1}$. The distance between $S_1$ and $S'$ is far enough but still finite in the thermodynamic limit. Hence we have $R \cap S'=\varnothing$ and $R_1 \cap R'=\varnothing$. In fact, this construction is also consistent with the other two cases.  In the second case, both two terms are unchanged after twisting. We can take $R_2=\varnothing$ and the sublattice $S'$ only need to be below the twisted boundary and thus satisfy $R \cap S'=\varnothing$. Thus both the $H_0$ and $H_{\text{dual}}$ are also unchanged after conjugating  $\prod_{v\in R'}X^{(a_1)}_v$. In the third case, the region of $H_{\text{dual}}$ belongs to $R_1$ and we also construct $R'$ so that it is below the twisted boundary ($R_1 \cap R'=\varnothing$) and  $R \cap S'=\varnothing$. Then $H_{\text{dual}}$ is unchanged after conjugating  $\prod_{v\in R'}X^{(a_1)}_v$.


 According to the discussion in the section \ref{2dproof} and \ref{anydproof}, we can obtain
 \begin{eqnarray}
  (\prod_{v\in R'}X^{(a_1)}_v)U_{(\underbrace{0,0,\cdots,0}_{d+1})}(\prod_{v\in R'}X^{(a_1)}_v)=U_{(\underbrace{0,0,\cdots,0}_{d+1})}U_{(\underbrace{0,0,\cdots,0}_{d})}({S_1\oplus S'}).
  \end{eqnarray}
  
   Thus, the dual term after twisting can be rewritten as 
\begin{eqnarray}
H^{\text{tw}}_{\text{dual}}&&=(\prod_{v\in R'}X^{(a_1)}_v)U_{(\underbrace{0,0,\cdots,0}_{d+1})}H_{0}U^{\dagger}_{(\underbrace{0,0,\cdots,0}_{d+1})}(\prod_{v\in R'}X^{(a_1)}_v)\nonumber\\&&=U_{(\underbrace{0,0,\cdots,0}_{d+1})}U_{(\underbrace{0,0,\cdots,0}_{d})}({S_1\oplus S'}) H^{\text{tw}}_0 U^{\dagger}_{(\underbrace{0,0,\cdots,0}_{d})}({S_1\oplus S'})U^{\dagger}_{(\underbrace{0,0,\cdots,0}_{d+1})}\nonumber\\&&=U_{(\underbrace{0,0,\cdots,0}_{d+1})}U_{(\underbrace{0,0,\cdots,0}_{d})}({S_1}) H^{\text{tw}}_0 U^{\dagger}_{(\underbrace{0,0,\cdots,0}_{d})}({S_1})U^{\dagger}_{(\underbrace{0,0,\cdots,0}_{d+1})},
\end{eqnarray}
where the last equality comes from $R\cap S'=\varnothing$.

Finally, we can conclude that  the twisted Hamiltonian possesses a modified duality symmetry 
\begin{eqnarray}
U^{(1)}_{(\underbrace{0,0,\cdots,0}_{d+1})}=U_{(\underbrace{0,0,\cdots,0}_{d+1})}U_{(\underbrace{0,0,\cdots,0}_{d})}(S_1).
\end{eqnarray}




\sloppy

\end{widetext}

\end{document}